\RequirePackage{fix-cm}
\documentclass[smallextended]{svjour3}       
\smartqed  
\usepackage{graphicx,amssymb,amsfonts,mathrsfs,latexsym,amsmath,amssymb}
\usepackage{color}
\journalname{}
\begin{document}

\title{Matrix-product structure of constacyclic codes over
finite chain rings $\mathbb{F}_{p^m}[u]/\langle u^e\rangle$
} \subtitle{}

\titlerunning{Matrix-product structure of constacyclic codes over
finite chain rings}

\author{Yuan Cao$^1$ $\cdot$  Yonglin Cao$^{1}$ $\cdot$ Fang-Wei Fu$^2$ 
}


\institute{Yonglin Cao (corresponding author) \at
             \email{ylcao@sdut.edu.cn}\\
           Yuan Cao\at
             \email{yuancao@sdut.edu.cn}\\
           Fang-Wei Fu\at
           fwfu@nankai.edu.cn\\
           $^1$School of Mathematics and Statistics, Shandong University of Technology, Zibo, Shandong 255091, China\\
           $^2$Chern Institute of Mathematics and LPMC, Nankai University, Tianjin 300071, China
          }

\date{Received: date / Accepted: date}

\maketitle

\begin{abstract}
Let $m,e$ be positive integers, $p$ a prime number,  $\mathbb{F}_{p^m}$ be a finite field of $p^m$ elements and $R=\mathbb{F}_{p^m}[u]/\langle u^e\rangle$ which is a finite chain ring.
For any $\omega\in R^\times$ and positive integers $k, n$ satisfying ${\rm gcd}(p,n)=1$, we prove that
any $(1+\omega u)$-constacyclic code of length $p^kn$ over $R$ is monomially equivalent to a matrix-product code of a nested sequence of $p^k$ cyclic codes with length $n$ over $R$ and a $p^k\times p^k$ matrix $A_{p^k}$ over $\mathbb{F}_p$. Using the matrix-product structures, we give an iterative construction of every $(1+\omega u)$-constacyclic code by $(1+\omega u)$-constacyclic codes of shorter lengths over $R$.

\keywords{Repeated-root constacyclic code \and Matrix-product code \and Monomially equivalent codes \and Finite chain ring
\vskip 3mm \noindent
{\bf Mathematics Subject Classification (2000)} 94B15 \and 94B05 \and 11T71}
\end{abstract}

\section{Introduction}
\noindent
  Algebraic coding theory deals with the design of error-correcting and error-detecting codes for the reliable transmission
of information across noisy channel. The class of constacyclic codes play a very significant role in
the theory of error-correcting codes.

\par
  Let $\Gamma$ be a commutative finite chain ring with identity $1\neq 0$, and $\Gamma^{\times}$ be the multiplicative group of invertible elements of
$\Gamma$. For any $a\in
\Gamma$, we denote by $\langle a\rangle_\Gamma$, or $\langle a\rangle$ for
simplicity, the ideal of $\Gamma$ generated by $a$, i.e. $\langle
a\rangle_\Gamma=a\Gamma=\{ab\mid b\in \Gamma\}$. For any ideal $I$ of $\Gamma$, we will identify the
element $a+I$ of the residue class ring $\Gamma/I$ with $a$ (mod $I$) for
any $a\in \Gamma$.

\par
   A \textit{code} of length $N$ over $\Gamma$ is a nonempty subset ${\cal C}$ of $\Gamma^N=\{(a_0,a_1,\ldots$, $a_{N-1})\mid a_j\in \Gamma, \ j=0,1,\ldots,N-1\}$. Each element of ${\cal C}$ is called a \textit{codeword} and the number of codewords in ${\cal C}$ is denoted by $|{\cal C}|$. The code ${\cal C}$
is said to be \textit{linear} if ${\cal C}$ is a $\Gamma$-submodule of $\Gamma^N$. For any codeword $c=(c_0,c_1,\ldots,c_{N-1})\in \mathcal{C}$, the \textit{Hamming weight} of $c$ is defined by ${\rm w}_H(c)=|\{j\mid c_j\neq 0, \
0\leq j\leq N-1\}|$. Then
the \textit{minimum Hamming distance} of a linear code ${\cal C}$ is equal to $d_H(\mathcal{C})={\rm min}\{{\rm w}_H(c)\mid c\neq 0, \ c\in \mathcal{C}\}$. If $M=|{\cal C}|$ and $d=d_H(\mathcal{C})$, ${\cal C}$ is called an $(N, M, d)$-code over $\Gamma$. All codes in this paper are assumed to be linear.

\par
   Let $\gamma\in \Gamma^{\times}$. A linear code
${\cal C}$ of length $N$ over $\Gamma$ is
called a $\gamma$-\textit{constacyclic code}
if $(\gamma c_{N-1},c_0,c_1,\ldots,c_{N-2})\in {\cal C}$ for all
$(c_0,c_1,\ldots,c_{N-1})\in{\cal C}$. Particularly, ${\cal C}$ is
called a \textit{negacyclic code} if $\gamma=-1$, and ${\cal C}$ is
called a  \textit{cyclic code} if $\gamma=1$.

\par
  For any $a=(a_0,a_1,\ldots,a_{N-1})\in \Gamma^N$, let
$a(x)=a_0+a_1x+\ldots+a_{N-1}x^{N-1}\in \Gamma[x]/\langle x^N-\gamma\rangle$. We will identify $a$ with $a(x)$ in
this paper. It is well known that ${\cal C}$ is a  $\gamma$-constacyclic code
of length $N$ over $\Gamma$ if and only if ${\cal C}$ is an ideal of
the residue class ring $\Gamma[x]/\langle x^N-\gamma\rangle$. Let $p$ be the characteristic of the residue class field of $\Gamma$. If ${\rm gcd}(p,N)=1$, ${\cal C}$ is called a
\textit{simple-root constacyclic code} while when $p\mid N$ it is called a \textit{repeated-root constacyclic code}.

\par
  For any positive integer $N$, we denote $[N)=\{0,1,\ldots,N-1\}$ in this paper. Let $C_1$ and $C_2$ be
codes of length $N$ over $\Gamma$. Recall that $C_1$ and $C_2$ are said to be \textit{monomially equivalent} if
there exists a permutation $\varrho$ on the set $[N)$ and fixed elements $r_0,r_1,\ldots,r_{N-1}\in \Gamma^\times$
such that
$$C_2=\{(r_0c_{\varrho(0)},r_1c_{\varrho(1)},\ldots,r_{N-1}c_{\varrho(N-1)})\mid (c_0,c_1,\ldots,c_{N-1})\in C_1\}$$
(cf. Huffman and  Pless [13] Page 24). Especially, $C_1$ and $C_2$ are said to be \textit{permutation equivalent} when $r_0=r_1=\ldots=r_{N-1}=1$ (cf. [13] Page 20). Recall that a \textit{monomial matrix} over $\Gamma$ is a square matrix with exactly one invertible entry
in each row and column. Hence $C_1$ and $C_2$ are monomially equivalent
if and only if there is an $N\times N$ monomial matrix $Q$ over $\Gamma$ such that
$Q\cdot C_1=\{Q\xi\mid \xi\in C_1\}=C_2$ in which we regard each $\xi\in C_1$ as an $N\times 1$ column vector over $\Gamma$.

\par
  From now on, let $m$ and $e$ be positive integers, $p$ a prime number,  $\mathbb{F}_{p^m}$ be a finite field of $p^m$ elements and denote $$R=\mathbb{F}_{p^m}[u]/\langle u^e\rangle=\mathbb{F}_{p^m}+u\mathbb{F}_{p^m}+\ldots+u^{e-1}\mathbb{F}_{p^m}
\ (u^e=0).$$
 It is known
that $R$ is a finite chain ring with subfield $\mathbb{F}_{p^m}$, $uR$ is the unique maximal ideal and $e$ is the nilpotency index of $u$.
All invertible elements of $R$ are given by
$a_0+a_1 u+\ldots+ a_{e-1} u^{e-1}, \ a_0\neq 0, \ a_0, a_1,\ldots,a_{e-1}\in \mathbb{F}_{p^m}.$

\par
There are many research results on constacyclic codes over $R$, see [1], [5--10] and [14] for examples.
Let $\omega\in R^\times$, $k$ and $n$ be positive integers satisfying ${\rm gcd}(p,n)=1$.
In this paper, we concentrate on $(1+\omega u)$-constacyclic codes of length $p^kn$ over $R$, i.e.
ideals of the residue class ring $R[x]/\langle x^{p^kn}-(1+\omega u)\rangle$. Specifically, the algebraic structures and properties of
$(1+w\gamma)$-constacyclic codes of arbitrary length over an arbitrary finite
chain ring $\Gamma$ were given in [4], where $w$ is a unit in $\Gamma$ and $\gamma$ generates the unique maximal ideal of $\Gamma$.

\par
  Blackford [2] classified all negacyclic codes over
the finite chain ring $\mathbb{Z}_4$ of even length
using a Discrete Fourier Transform approach. Using the concatenated structure given by [2] Theorem 3, we know that
each negacyclic code of length $2^kn$, where $n$ is odd, is monomially equivalent to a sequence of $2^k$ cyclic codes
of length $n$ over $\mathbb{Z}_4$.

\par
  As $-1=1+2\in \mathbb{Z}_4$, negacyclic codes of even length over
$\mathbb{Z}_4$ is a special subclass of the class of $(1+w\gamma)$-constacyclic codes with arbitrary length over an arbitrary finite
chain ring $\Gamma$. Now,
we try to give a matrix-product
structure for any $(1+\omega u)$-constacyclic code of length $p^kn$ over $R$ by us of the theory of finite chain rings.
    In this paper, we denote
$$\mathcal{R}_k:=R[v]/\langle v^{p^k}-(1+\omega u)\rangle.$$
As $R[x]/\langle f\rangle=R$ when $f=x-1$, from Cao [4] Theorem 2.4 and Dinh et al [10] Section 4
we deduce the following lemma.

\vskip 3mm \noindent
  {\bf Lemma 1.1} \textit{Using the notations above, we have the following conclusions}.

\par
  (i) \textit{$v-1$ is nilpotent in the ring $\mathcal{R}_k$}.

\par
  (ii) \textit{$\mathcal{R}_k$ is a commutative finite chain ring with maximal ideal $(v-1)\mathcal{R}_k$,
and $p^ke$ is the nilpotency index of $v-1$. Furthermore, $u\mathcal{R}_k=(v-1)^{p^k}\mathcal{R}_k$}.

\par
  (iii) \textit{$\mathcal{R}_k/(v-1)\mathcal{R}_k\cong \mathbb{F}_{p^m}$}.

\par
  (iv) \textit{All $p^{k}e+1$ distinct ideals of $\mathcal{R}_k$ are given by}
$$\{0\}=(v-1)^{p^{k}e}\mathcal{R}_k\subset (v-1)^{p^ke-1}\mathcal{R}_k\subset\ldots\subset(v-1)\mathcal{R}_k\subset(v-1)^0\mathcal{R}_k=\mathcal{R}_k.$$
\textit{Moreover, the number of elements in $(v-1)^i\mathcal{R}_k$ is equal to $|(v-1)^i\mathcal{R}_k|=p^{m(p^{k}e-i)}$ for all
$i=0,1,\ldots,p^{k}e$}.

\vskip 3mm\par
  We will construct a precise isomorphism of rings from $R[x]/\langle x^{p^kn}-(1+\omega u)\rangle$
onto $\mathcal{R}_k[x]/\langle x^n-1\rangle$, which induces a one-to-one correspondence between the set
of  $(1+\omega u)$-constacyclic codes of length $p^kn$ over
$R$ onto the set of cyclic codes of length $n$ over
$\mathcal{R}_k$.
   By the theory of simple-root cyclic codes over finite chain rings (cf. Norton et al [15]), any cyclic code of length $n$ over
$\mathcal{R}_k$ can be determined uniquely by a tower of $p^ke$ cyclic
codes with length $n$ over the finite field $\mathbb{F}_{p^m}$
$$\langle g_0(x)\rangle\subseteq \langle g_1(x)\rangle\subseteq\ldots\subseteq \langle g_{p^{k}e-1}(x)\rangle\subseteq \mathbb{F}_{p^m}[x]/\langle x^n-1\rangle,$$
where $g_0(x),g_1(x), \ldots, g_{p^{k}e-1}(x)$ are monic divisors of $x^n-1$ in $\mathbb{F}_{p^m}[x]$ satisfying
$g_{p^{k}e-1}(x)\mid \ldots\mid g_1(x)\mid g_0(x)\mid (x^n-1)$.
Then we give a direct description of a monomially equivalence between
a $(1+\omega u)$-constacyclic code of length $p^kn$ over $R$ and a matrix-product code of a sequence of $p^k$ cyclic codes
over $R$ determined by $g_s(x)$, $s=0,1,\ldots,p^{k}e-1$.

\par
   In Section 2, we sketch the concept of matrix-product codes and structures of simple-root cyclic codes over the finite chain ring $\mathcal{R}_k$. In Section 3, we prove that any $(1+\omega u)$-constacyclic code of length $p^kn$ over $R$ is monomially equivalent
to a matrix-product code of a nested sequence of $p^{k}$ cyclic codes with length $n$ over $R$. Using this matrix-product structure, we give an iterative construction of every $(1+\omega u)$-constacyclic code by use of $(1+\omega u)$-constacyclic codes of shorter lengths over $R$ in Section 4. In Section 5,
we consider how to get the matrix-product structures of $(1+u)$-constacyclic codes of length $90$ over $R=\mathbb{F}_3+u\mathbb{F}_3$ ($u^2=0$).



\section{Preliminaries}
\noindent
   In this section, we sketch the concept of matrix-product codes and structures of simple-root cyclic codes over the finite chain ring $\mathcal{R}_k$.

  Let $R=\mathbb{F}_{p^m}[u]/\langle u^e\rangle$. We follow the notation in [3] Definition 2.1 for definition of matrix-product codes.
Let $A=[a_{ij}]$ be an $\alpha\times\beta$ matrix with entries in $R$ and let
$C_1,\ldots,C_\alpha$ be codes of length $n$ over $R$. The \textit{matrix-product code}
$[C_1,\ldots,C_\alpha]\cdot A$ is the set of all matrix products $[c_1,\ldots,c_\alpha]\cdot A$ defined by
\begin{eqnarray*}
[c_1,\ldots,c_\alpha]\cdot A&=&[c_1,\ldots,c_\alpha]\left[\begin{array}{cccc}a_{11} & a_{12} & \ldots & a_{1\beta}\cr
a_{21} & a_{22} & \ldots & a_{2\beta}\cr \vdots &\vdots &\vdots &\vdots \cr a_{\alpha 1} & a_{\alpha 2} & \ldots & a_{\alpha \beta}\end{array}\right]\\
 &=& [a_{11}c_1+a_{21}c_2+\ldots+a_{\alpha 1}c_{\alpha}, a_{12}c_1+a_{22}c_2+\ldots+a_{\alpha 2}c_{\alpha},\\
    && \ldots, a_{1\beta}c_1+a_{2\beta}c_2+\ldots+a_{\alpha \beta}c_{\alpha}]
\end{eqnarray*}
where $c_i\in C_i$ is an $n\times 1$ column vector for $1\leq i\leq \alpha$. Any codeword $[c_1,\ldots,c_\alpha]\cdot A$
is an $n\times \beta$ matrix over $R$ and we regard it as a codeword of length $n\beta$ by reading the entries of
the matrix in column-major order.
A code $C$ over $R$
is a matrix-product code if $C=[C_1,\ldots,C_\alpha]\cdot A$ for some codes
$C_1,\ldots,C_\alpha$ and a matrix $A$.

\par
  In the rest of this paper, we assume that $A=[a_{ij}]$ is an $\alpha\times\beta$ matrix over $\mathbb{F}_{p^m}$, i.e. $a_{ij}\in \mathbb{F}_{p^m}$ for all $i,j$. If the rows of $A$ are linearly independent over $\mathbb{F}_{p^m}$, $A$ is called a \textit{full-row-rank} (FRR) matrix.
Let $A_t$ be the matrix consisting of the first $t$ rows of $A$.
For $1\leq j_1 < j_2 < \ldots < j_t \leq \beta$, we denote by $A(j_1, j_2, \ldots, j_t)$ the $t\times t$ submatrix consisting
of the columns $j_1, j_2, \ldots, j_t$ of $A_t$. If every sub-matrix $A(j_1, j_2, \ldots, j_t)$ of $A$ is non-singular
for all $t=1,\ldots,\alpha$, $A$ is said to be \textit{non-singular by columns} (NSC) (cf. [3] Definition 3.1).

\par
  As a natural generalization of [12] Theorem 1 and results in [16], by [11] Theorem 3.1 we have the following
properties of matrix-product codes.

\vskip 3mm \noindent
  {\bf Theorem 2.1} \textit{Let $A$ be an $\alpha\times \beta$ FRR matrix over $\mathbb{F}_{p^m}$, and
$C_i$ be a linear $(n, M_i, d_i)$-code over $R$ for all $i=1,\ldots,\alpha$. Then the matrix-product code $[C_1, ..., C_\alpha] \cdot A$  is a linear $(n\beta, \prod_{i=1}^\alpha M_i, d)$-code over $R$ where the minimum Hamming distance $d$ satisfies
$$d\geq\delta:={\rm min}\{\delta_id_i\mid i=1,\ldots,\alpha\},$$
where $\delta_i$ is the minimum distance of the linear code with length $\beta$ over $\mathbb{F}_{p^m}$ generated by the first $i$ rows of
the matrix $A$}.

\par
   \textit{Moreover, when the matrix $A$ is NSC, it holds that
$\delta_i=\beta-i+1$. Furthermore, if we assume that the codes $C_i$ form a nested sequence
$C_1 \supseteq C_2 \supseteq \ldots\supseteq C_\alpha$, then $d=\delta$}.

\vskip 3mm
\par
   Then we consider cyclic codes
of length $n$ over the finite chain ring $\mathcal{R}_k=R[v]/\langle v^{p^k}-(1+\omega u)\rangle$, i.e. ideals of the residue class ring
$\mathcal{R}_k[x]/\langle x^n-1\rangle$. Let $\alpha\in \mathcal{R}_k$. By
Lemma 1.1 and properties of finite chain rings, $\alpha$ has a unique $(v-1)$-expansion
$$\alpha=\sum_{s=0}^{p^{k}e-1}a_s(v-1)^s, \ a_s\in \mathbb{F}_{p^m}, \ s=0,1,\ldots, p^{k}e-1.$$
In this paper, we define $\tau: \mathcal{R}_k\rightarrow \mathbb{F}_{p^m}$ by
$$\tau(\alpha)=a_0=\alpha \ ({\rm mod} \ v-1), \  \forall \alpha\in \mathcal{R}_k.$$
Then $\tau$ is a surjective homomorphism of rings from $\mathcal{R}_k$ onto $\mathbb{F}_{p^m}$.
As $u\mathcal{R}_k=(v-1)^{p^k}\mathcal{R}_k$ by Lemma 1.1(ii), there is an invertible element $\varepsilon\in \mathcal{R}_k^\times$  such that $u=(v-1)^{p^k}\varepsilon$, which implies $\tau(u)=0$. Hence for any $\beta=b_0+b_1u+\ldots+b_{e-1}u^{e-1}\in R\subseteq \mathcal{R}_k$
where $b_0,b_1,\ldots,b_{e-1}\in \mathbb{F}_{p^m}$, we have
\begin{equation}
\tau(\beta)=b_0=\beta \ ({\rm mod} \ u).
\end{equation}
It is clear that $\tau$ can be extended to
a surjective homomorphism of polynomial rings from $\mathcal{R}_k[x]$ onto $\mathbb{F}_{p^m}[x]$ by:
$\sum\alpha_ix^i\mapsto \sum\tau(\alpha_i)x^i, \ \forall \alpha_i\in \mathcal{R}_k.$
We still use $\tau$ to denote this homomorphism. Then $\tau$ induces a surjective homomorphism of rings from
$\mathcal{R}_k[x]/\langle x^n-1\rangle$ onto $\mathbb{F}_{p^m}[x]/\langle x^n-1\rangle$
in the natural way
$$\tau(\sum_{i=0}^{n-1}\alpha_ix^i)=\sum_{i=0}^{n-1}\tau(\alpha_i)x^i, \ \forall \alpha_0,\alpha_1,\ldots,\alpha_{n-1}\in \mathcal{R}_k.$$

\par
   Now, let $\mathcal{C}$ be a cyclic code
of length $n$ over $\mathcal{R}_k$. For any integer $s$, $0\leq s\leq p^{k}e-1$, define
$$(\mathcal{C}:(v-1)^s)=\left\{\alpha(x)\in \mathcal{R}_k[x]/\langle x^n-1\rangle\mid (v-1)^s\alpha(x)\in \mathcal{C}\right\}$$
which is an ideal of $\mathcal{R}_k[x]/\langle x^n-1\rangle$ as well. It is clear that
\begin{equation}
\mathcal{C}=(\mathcal{C}:(v-1)^0)\subseteq (\mathcal{C}:(v-1))\subseteq \ldots (\mathcal{C}:(v-1)^{p^{k}e-1}).
\end{equation}
Denote
$${\rm Tor}_s(\mathcal{C})=\tau(\mathcal{C}:(v-1)^s)=\{\tau(\alpha(x))\mid \alpha(x)\in (\mathcal{C}:(v-1)^s)\}.$$
Then ${\rm Tor}_s(\mathcal{C})$
is an ideal of the ring $\mathbb{F}_{p^m}[x]/\langle x^n-1\rangle$, i.e. a cyclic code
of length $n$ over $\mathbb{F}_{p^m}$, which is called the \textit{$s$th torsion code} of $\mathcal{C}$. Hence there is a unique
monic divisor $g_s(x)$ of $x^n-1$ in $\mathbb{F}_{p^m}[x]$ such that
$${\rm Tor}_s(\mathcal{C})=\langle g_s(x)\rangle=\{b(x)g_s(x)\mid {\rm deg}(b(x))<n-{\rm deg}(g_s(x)), \ b(x)\in \mathbb{F}_{p^m}[x]\},$$
where $g_s(x)$ is the generator polynomial of the cyclic
code ${\rm Tor}_s(\mathcal{C})$. Hence $|{\rm Tor}_s(\mathcal{C})|=p^{m(n-{\rm deg}(g_s(x)))}$.

\par
  As $g_s(x)\in {\rm Tor}_s(\mathcal{C})$, we have $(v-1)^s(g_s(x)-(v-1) b_s(x))\in \mathcal{C}$ for some $b_s(x)\in \mathcal{R}_k[x]$.
Then by $(v-1)^{p^{k}e}=0$ in $\mathcal{R}_k$, it follows that
\begin{eqnarray*}
(v-1)^sg_s(x)^{p^{k}e-s}
&=&(v-1)^s\left(g_s(x)^{p^{k}e-s}-(v-1)^{p^{k}e-s}b_s(x)^{p^{k}e-s}\right)\\
&=&(v-1)^s(g_s(x)-(v-1) b_s(x))\\
 &&\cdot \left(\sum_{t=0}^{p^{k}e-s-1}g_s(x)^t\cdot\left((v-1)b_s(x)\right)^{p^{k}e-s-1-t}\right).
\end{eqnarray*}
This implies $(v-1)^sg_s(x)^{p^{k}e-s}\in \mathcal{C}$. As ${\rm gcd}(p,n)=1$, $x^n-1$ has no repeated divisors in $\mathbb{F}_{p^m}[x]$. This implies
${\rm gcd}(x^n-1,g_s(x)^{p^{k}e-s})=g_s(x)$. Hence there exist $a(x),b(x)\in \mathbb{F}_{p^m}[x]$ such
that $g_s(x)=a(x)g_s(x)^{p^{k}e-s}+b(x)(x^n-1)=a(x)g_s(x)^{p^{k}e-s}$ in $\mathcal{R}_k[x]/\langle x^n-1\rangle$. Therefore, we have
\begin{equation}
(v-1)^sg_s(x)=a(x)\cdot (v-1)^sg_s(x)^{p^{k}e-s}\in \mathcal{C}, \ s=0,1,\ldots,p^{k}e-1.
\end{equation}
This implies $(v-1)^s{\rm Tor}_s(\mathcal{C})\subseteq \mathcal{C}$ for all $ s=0,1,\ldots,p^{k}e-1$. Moreover, by Equation (2) we have a tower of cyclic
codes over $\mathbb{F}_{p^m}$:
$${\rm Tor}_0(\mathcal{C})\subseteq {\rm Tor}_1(\mathcal{C})\subseteq\ldots\subseteq {\rm Tor}_{p^{k}e-1}(\mathcal{C})\subseteq \mathbb{F}_{p^m}[x]/\langle x^n-1\rangle.$$
This implies that
$g_{p^{k}e-1}(x)\mid\ldots \mid g_1(x)\mid g_0(x)\mid(x^n-1)$ in $\mathbb{F}_{p^m}[x]$.

\par
   Now, let $c(x)\in \mathcal{C}$. Then $\tau(c(x))\in {\rm Tor}_0(\mathcal{C})=\langle g_0(x)\rangle$. Hence there exists a unique polynomial $b_0(x)\in \mathbb{F}_{p^m}[x]$ satisfying ${\rm deg}(b_0(x))<n-{\rm deg}(g_0(x))$ such that $\tau(c(x))=b_0(x)g_0(x)$.
By Equation (3), it follows that $b_0(x)g_0(x)\in \mathcal{C}$. Hence
$c(x)-b_0(x)g_0(x)\in \mathcal{C}$.

\par
  As $\tau(c(x)-b_0(x)g_0(x))=\tau(c(x))-b_0(x)g_0(x)=0$, there exists $\alpha_1(x)\in \mathcal{R}_k[x]/\langle x^n-1\rangle$ such that $(v-1)\alpha_1(x)=c(x)-b_0(x)g_0(x)\in \mathcal{C}$. This implies
$\alpha_1(x)\in (\mathcal{C}:(v-1))$, and so $\tau(\alpha_1(x))\in {\rm Tor}_1(\mathcal{C})$.

\par
   By ${\rm Tor}_1(\mathcal{C})=\langle g_1(x)\rangle$, there exists a unique polynomial $b_1(x)\in \mathbb{F}_{p^m}[x]$ satisfying ${\rm deg}(b_1(x))<n-{\rm deg}(g_1(x))$ such that $\tau(\alpha_1(x))=b_1(x)g_1(x)$. Then by Equation (3), it follows that
$(v-1)b_1(x)g_1(x)=b_1(x)\cdot (v-1)g_1(x)\in \mathcal{C}$. By $\tau(\alpha_1(x)-b_1(x)g_1(x))=0$, there exists $\alpha_2(x)\in \mathcal{R}_k[x]/\langle x^n-1\rangle$ such that $(v-1)\alpha_2(x)=\alpha_1(x)-b_1(x)g_1(x)$ and
$$(v-1)^2\alpha_2(x)=(v-1)\alpha_1(x)-(v-1)b_1(x)g_1(x)\in \mathcal{C}.$$
This implies
$\alpha_2(x)\in (\mathcal{C}:(v-1)^2)$, and so $\tau(\alpha_2(x))\in {\rm Tor}_2(\mathcal{C})=\langle g_2(x)\rangle$.

\par
  As stated above, we have
\begin{eqnarray*}
c(x)&=&b_0(x)g_0(x)+(v-1)\alpha_1(x)\\
 &=&b_0(x)g_0(x)+(v-1)b_1(x)g_1(x)+(v-1)^2\alpha_2(x),
\end{eqnarray*}
where $c_0(x)=b_0(x)g_0(x)\in {\rm Tor}_0(\mathcal{C})$ and $c_1(x)=b_1(x)g_1(x)\in {\rm Tor}_1(\mathcal{C})$.

\par
  Let $2\leq s\leq p^{k}e-2$ and assume that there exist $c_i(x)\in {\rm Tor}_i(\mathcal{C})$, $i=0,1,\ldots,s$, and
$\alpha_{s+1}(x)\in \mathcal{R}_k[x]/\langle x^n-1\rangle$ such that
$$c(x)=\sum_{i=0}^s(v-1)^ic_i(x)+(v-1)^{s+1}\alpha_{s+1}(x).$$
Then by $(v-1)^ic_i(x)\in (v-1)^i{\rm Tor}_i(\mathcal{C})\subseteq \mathcal{C}$, it follows that
$(v-1)^{s+1}\alpha_{s+1}(x)\in \mathcal{C}$. This implies $\alpha_{s+1}(x)\in (\mathcal{C}:(v-1)^{s+1})$, and so
$\tau(\alpha_{s+1}(x))\in {\rm Tor}_{s+1}(\mathcal{C})=\langle g_{s+1}(x)\rangle$. We denote
$c_{s+1}(x)=\tau(\alpha_{s+1}(x))$. Then there exists $\alpha_{s+2}(x)\in \mathcal{R}_k[x]/\langle x^n-1\rangle$ such that
$\alpha_{s+1}(x)=c_{s+1}(x)+(v-1)\alpha_{s+2}(x)$, and hence
$(v-1)^{s+1}\alpha_{s+1}(x)=(v-1)^{s+1}c_{s+1}(x)+(v-1)^{s+2}\alpha_{s+2}(x)$. Therefore,
$$c(x)=\sum_{i=0}^{s+1}(v-1)^ic_i(x)+(v-1)^{s+2}\alpha_{s+2}(x).$$
By mathematical induction on $s$, we conclude the following theorem.

\vskip 3mm \noindent
  {\bf Theorem 2.2}  \textit{Using the notations above, we have the following conclusions}.

\par
 (i) \textit{Let $\mathcal{C}$ be a cyclic code of length $n$ over $\mathcal{R}_k=R[v]/\langle v^{p^k}-(1+\omega u)\rangle$. Then each
codeword $c(x)$ in $\mathcal{C}$ has a unique $(v-1)$-adic expansion}:
$$c(x)=\sum_{s=0}^{p^{k}e-1}(v-1)^sc_s(x), \ {\rm where} \ c_s(x)\in {\rm Tor}_s(\mathcal{C}), \ \forall s=0,1,\ldots,p^{k}e-1.$$
\textit{Hence $|\mathcal{C}|=\prod_{s=0}^{p^{k}e-1}|{\rm Tor}_s(\mathcal{C})|=p^{m(\sum_{s=0}^{p^{k}e-1}(n-{\rm deg}(g_s(x))))}$}.

\par
 (ii) \textit{$\mathcal{C}$ is a cyclic code of length $n$ over $\mathcal{R}_k$ if and only if there exists uniquely
a tower of $p^ke$ cyclic
codes with length $n$ over $\mathbb{F}_{p^m}$,
$C_0\subseteq C_1\subseteq\ldots\subseteq C_{p^{k}e-1},$
such that ${\rm Tor}_s(\mathcal{C})=\tau(\mathcal{C}:(v-1)^s)=C_s$ for all $s=0,1,\ldots,p^{k}e-1$. If the latter conditions are satisfied, then
\begin{eqnarray*}
\mathcal{C}&=&\bigoplus_{s=0}^{p^ke-1}(v-1)^sC_s\\
 &=&\left\langle g_0(x),(v-1)g_1(x),\ldots,(v-1)^{p^ke}g_{p^ke-1}(x)\right\rangle_{\mathcal{R}_k[x]/\langle x^n-1\rangle}\\
 &=&\left\langle \sum_{s=0}^{p^{k}e-1}(v-1)^sg_s(x)\right\rangle_{\mathcal{R}_k[x]/\langle x^n-1\rangle}
\end{eqnarray*}
where $g_s(x)\in \mathbb{F}_{p^m}[x]$ being the generator polynomial of the cyclic code $C_s$ for all $s=0,1,\ldots,p^{k}e-1$}.

\vskip 3mm \noindent
 \textit{Remark} For a complete description of simple-root cyclic codes over arbitrary commutative finite chain rings,
readers can refer to [15] Theorem 3.5.

\vskip 3mm\par
   When $k=0$, we have $\mathcal{R}_0=R[v]/\langle v-(1+\omega u)\rangle=R$ satisfying $v-1=\omega u$ or $u=\omega^{-1}(v-1)$. Then from Lemma 1.1,
Theorem 2.2 and Equation (1), we deduce the following corollary which will be used in the following sections.

\vskip 3mm \noindent
 {\bf Corollary 2.3} \textit{Using the notations above, we have the following conclusions}.

\par
 (i) \textit{Let $\mathcal{C}$ be a cyclic code of length $n$ over $R=\mathbb{F}_{p^m}[u]/\langle u^e\rangle$. Then each
codeword $c(x)$ in $\mathcal{C}$ has a unique $u$-adic expansion}:
$$c(x)=\sum_{s=0}^{e-1}u^sc_s(x), \ {\rm where} \ c_s(x)\in {\rm Tor}_s(\mathcal{C})=\tau(\mathcal{C}:u^s), \ \forall s=0,1,\ldots,e-1.$$
\textit{Hence $|\mathcal{C}|=\prod_{s=0}^{e-1}|{\rm Tor}_s(\mathcal{C})|=p^{m(\sum_{s=0}^{e-1}(n-{\rm deg}(g_s(x))))}$}.

\par
 (ii) \textit{$\mathcal{C}$ is a cyclic code of length $n$ over $R$ if and only if there exists uniquely
a tower of $e$ cyclic
codes with length $n$ over $\mathbb{F}_{p^m}$,
$C_0\subseteq C_1\subseteq\ldots\subseteq C_{e-1},$
such that ${\rm Tor}_s(\mathcal{C})=C_s$ for all $s=0,1,\ldots,e-1$. If the latter conditions are satisfied, then
$\mathcal{C}=\bigoplus_{s=0}^{e-1}u^sC_s$ and
$|\mathcal{C}|=\prod_{i=0}^{e-1}|C_i|$. Furthermore, we have
$$\mathcal{C}=\left\langle g_0(x),ug_1(x),\ldots,u^{e-1}g_{e-1}(x)\right\rangle_{R[x]/\langle x^n-1\rangle}
=\left\langle \sum_{s=0}^{e-1}u^sg_s(x)\right\rangle_{R[x]/\langle x^n-1\rangle}$$
where $g_s(x)\in \mathbb{F}_{p^m}[x]$ being the generator polynomial of the cyclic code $C_s$ for all $s=0,1,\ldots,e-1$}.

\par
 (iii) \textit{Let $\mathcal{C}$ and $\mathcal{C}^\prime$ be cyclic codes of length $n$ over $R$ with
$C_s={\rm Tor}_s(\mathcal{C})$ and $C_s^\prime={\rm Tor}_s(\mathcal{C}^\prime)$ for all $s$. Then
$\mathcal{C}\subseteq \mathcal{C}^\prime$ if and only if $C_s\subseteq C_s^\prime$ as ideals of the ring $\mathbb{F}_{p^m}[x]/\langle x^n-1\rangle$
 for all $s=0,1,\ldots,e-1$}.

\vskip 3mm\noindent
  \textit{Proof} We only need to prove (iii). If $C_s\subseteq C_s^\prime$ for all $s=0,1,\ldots,e-1$, it is
obvious that $\mathcal{C}=\bigoplus_{s=0}^{e-1}u^sC_s\subseteq \bigoplus_{s=0}^{e-1}u^sC_s^\prime=\mathcal{C}^\prime$.

\par
   Conversely, let
$\mathcal{C}\subseteq \mathcal{C}^\prime$. Then $(\mathcal{C}:u^s)\subseteq (\mathcal{C}^\prime:u^s)$ for all $s$. From this, by
${\rm Tor}_s(\mathcal{C})=\tau(\mathcal{C}:u^s)$ and ${\rm Tor}_s(\mathcal{C}^\prime)=\tau(\mathcal{C}^\prime:u^s)$ we deduce that
$C_s\subseteq C_s^\prime$ for all $s=0,1,\ldots,e-1$.
\hfill
$\Box$



\section{Matrix-product structure of $(1+\omega u)$-constacyclic codes over $R$}
\noindent
  Denote $[n)\times [p^k)=\{(j,t)\mid j\in [n), \ t\in [p^k)\}$. Then each integer $i\in [p^kn)=\{0,1,\ldots,p^kn-1\}$ can be uniquely expressed as
\begin{equation}
i=j+tn, \ {\rm were} \ j\equiv i \ ({\rm mod} \ n), \ j\in [n), \ {\rm and} \ t=\frac{i-j}{n}\in [p^k).
\end{equation}
In this paper, we adopt the following notations.

\vskip 3mm \noindent
  {\bf Notation 3.1} Let $l$ be the smallest positive integer such that $p^l\geq e$.
Since ${\rm gcd}(p,n)=1$, there exists a unique integer $n^\prime$,
$1\leq n^\prime \leq p^{k+l}-1$, such that
\begin{equation}
n^\prime n\equiv 1 \ ({\rm mod} \ p^{k+l}).
\end{equation}
We write $n^\prime=qp^k+n^{\prime\prime}$, where $0\leq q\leq p^l-1$
and $1\leq n^{\prime\prime}\leq p^k-1$ satisfying ${\rm gcd}(p,n^{\prime\prime})=1$. Then we
define a transformation $\varrho$ on the set $[p^kn)$ by
$$\varrho(j+\lambda n)=j+n\left(\lambda-jn^{\prime\prime}
\ ({\rm mod} \ p^k)\right), \ \forall (j,\lambda)\in [n)\times [p^k),$$
and denote
$$\Lambda={\rm diag}[1,(1+\omega u)^q,(1+\omega u)^{2q},\ldots,(1+\omega u)^{(n-1)q}]$$
which is a diagonal matrix of order $n$ with $1,(1+\omega u)^q,(1+\omega u)^{2q},\ldots,(1+\omega u)^{(n-1)q}\in R^\times$ as its diagonal entries.

\vskip 3mm \noindent
  {\bf Lemma 3.2}
  (i) \textit{The transformation $\varrho$ is a permutation on the set $[p^kn)$}.

\par
  (ii) \textit{Let $P_{p^kn}$ be a matrix of order $p^kn$ defined by
$P_{p^kn}=[\epsilon_{i,j}]$ where}
$$\epsilon_{i,j}=1 \ {\rm if} \ j=\varrho(i), \ {\textit and} \ \epsilon_{i,j}=0 \ {\textit othwise},
\ {\textit for} \ {\textit all} \ 0\leq i,j\leq p^kn-1,$$
\textit{and set $M_{p^k}(n,\omega)={\rm diag}[\stackrel{(p^k)^{,}{\rm s}}{\overbrace{\Lambda,\ldots,\Lambda}}]\cdot P_{p^kn}$. Then $P_{p^kn}$ is a permutation matrix
and $M_{p^k}(n,\omega)$ is a monomial matrix over $R$ of order $p^kn$}.

\par
  (iii) \textit{Define a transformation $\Theta$ on the $R$-module $R^{p^kn}$ by}
$$\Theta(\xi)=M_{p^k}(n,\omega)\cdot \xi, \ \forall \xi=\left[\begin{array}{c}a_0\cr a_1\cr \ldots \cr a_{p^kn-1}\end{array}\right]\in R^{p^kn}.$$
\textit{Then $\Theta$ is an $R$-module automorphism on $R^{p^kn}$. Let $C$ be an
$R$-submodule of $R^{p^kn}$ and denote $\Theta(C)=M_{p^k}(n,\omega)\cdot C=\{M_{p^k}(n,\omega)\cdot c\mid c\in C\}$. Then
$\Theta(C)$ and $C$ are  monomially equivalent linear codes of length $p^kn$ over $R$}.

\vskip 3mm \noindent
  \textit{Proof} (i) For any $(j,\lambda)\in [n)\times[p^k)$, let $t=\lambda-jn^{\prime\prime}$
(mod $p^k$). Then by Equation (4) and $(j,t)=(j,\lambda)\left[\begin{array}{cc}1 & -n^{\prime\prime}
\cr 0 & 1\end{array}\right]$, we see that $\varrho: j+\lambda n\mapsto j+tn$ ($\forall (j,\lambda)\in [n)\times[p^k)$) is a a permutation on the set $[p^kn)$.

\par
  (ii) follows from (i) and Notation 3.1, and (iii) follows from (ii).
\hfill $\Box$

\vskip 3mm\par
  First, we establish an explicit relationship between the set of all
$(1+\omega u)$-constacyclic codes of length $p^kn$ over the finite chain ring $R=\mathbb{F}_{p^m}[u]/\langle u^e\rangle$ and
the set of all cyclic codes of length $n$ over the finite chain ring $\mathcal{R}_k$.

\par
  Let $a(x)\in R[x]/\langle x^{p^kn}-(1+\omega u)\rangle$. By Equation (4), $a(x)$ can be uniquely expressed as
$a(x)=\sum_{j=0}^{n-1}\sum_{t=0}^{p^k-1}a_{j+tn}x^{j+tn}$, where $a_0,a_1,\ldots,a_{p^kn-1}\in R$. We will identify $a(x)$ with the column vector $[a_0,a_1,\ldots,a_{p^kn-1}]^{{\rm tr}}\in R^{p^kn}$ in this paper. By $x^{j+tn}=x^j(x^n)^t$, we can write $a(x)$ as a product of matrices
\begin{equation}
a(x)=[1,x,x^{2},\ldots,x^{n-1}]M_{a(x)}X
\end{equation}
where $X=[1,x^n,x^{2n},\ldots,x^{(p^k-1)n}]^{{\rm tr}}$ is the transpose of the $1\times p^k$ matrix
$[1,x^n,x^{2n},\ldots,x^{(p^k-1)n}]$ and
$M_{a(x)}=\left[\begin{array}{cccc}a_{0} & a_{0+n} & \ldots & a_{0+(p^k-1)n} \cr
 a_{1} & a_{1+n} & \ldots & a_{1+(p^k-1)n} \cr \ldots  & \ldots  & \ldots & \ldots  \cr
 a_{n-1} & a_{n-1+n} & \ldots & a_{n-1+(p^k-1)n}
\end{array}\right]$.

\par
  Set $v=x^n$ in  Equation (6). We obtain
$$a(x)=[1,x,x^{2},\ldots,x^{n-1}]M_{a(x)}V.$$
where $V=[1,v,v^2,\ldots,v^{p^k-1}]^{{\rm tr}}$ is the transpose of the $1\times p^k$ matrix
$[1,v,v^2$, $\ldots,v^{p^k-1}]$.
  We define a map
$\varphi: R[x]/\langle x^{p^kn}-(1+\omega u)\rangle\rightarrow \mathcal{R}_k/\langle x^n-v\rangle$
by
$$\varphi(a(x))=[1,x,x^{2},\ldots,x^{n-1}]\left(M_{a(x)}V\right)=\alpha_0+\alpha_1 x+\ldots+\alpha_{n-1} x^{n-1}$$
where $[\alpha_0,\alpha_1,\ldots,\alpha_{n-1}]=\left(M_{a(x)}V\right)^{{\rm tr}}\in \mathcal{R}_k^{n}$. Then from $\mathcal{R}_k=R[v]\langle v^{p^k}-(1+\omega u)\rangle$ and
$R[x]/\langle x^{p^kn}-(1+\omega u)\rangle=R[x,v]/\langle v^{p^k}-(1+\omega u),x^n-v\rangle$
as residue class rings, we
deduce the following conclusion.

\vskip 3mm \noindent
  {\bf Lemma 3.3} \textit{The map $\varphi$ is an isomorphism of rings from $R[x]/\langle x^{p^kn}-(1+\omega u)\rangle$
onto $\mathcal{R}_k/\langle x^n-v\rangle$}.

\vskip 3mm \par
   By Notation 3.1, we have $p^l\geq e$. From this, by $v^{p^k}=x^{p^kn}=1+\omega u$ and $u^e=0$ in $\mathcal{R}_k$ we deduce that
$$v^{p^{k+l}}=(1+\omega u)^{p^l}=1+\omega^{p^l} u^{p^l}=1+\omega^{p^l} u^{p^l-e}u^e=1.$$
Then by Equation (5), it follows that $(v^{n^\prime})^n=v^{n^\prime n}=v$. Now,
we define an automorphism of the polynomial ring $\mathcal{R}_k[x]$ by
$\psi(\beta(x))=\beta(v^{n^\prime}x)$ ($\forall \beta(x)\in \mathcal{R}_k[x]$).
Since $\psi(x^n-v)=(v^{n^\prime}x)^n-v=v(x^n-1)$ and $v\in\mathcal{R}_k^\times$, $\psi$
induces an ring isomorphism of residue class rings from $\mathcal{R}_k[x]/\langle x^n-v\rangle$
onto $\mathcal{R}_k[x]/\langle x^n-1\rangle$:
$$\alpha(x)\mapsto\alpha(v^{n^\prime}x)=[1,x,\ldots,x^{n-1}]{\rm diag}(1,v^{n^\prime},\ldots, (v^{n^\prime})^{n-1})[\alpha_0,\alpha_1,\ldots,\alpha_{n-1}]^{{\rm tr}}$$
for any $\alpha(x)=\alpha_0+\alpha_1 x+\ldots+\alpha_{n-1} x^{n-1}\in \mathcal{R}_k[x]/\langle x^n-v\rangle$.
We will still use $\psi$ to denote this ring isomorphism.
Hence $\psi(\alpha(x))=\alpha(v^{n^\prime}x)$ for all $\alpha(x)\in \mathcal{R}_k[x]/\langle x^n-v\rangle$.
  Then by Lemma 3.3, we conclude the following conclusion.

\vskip 3mm \noindent
  {\bf Lemma 3.4} \textit{Using the notations above, the map $\psi\varphi$ define by
$$\psi\varphi(a(x))=[1,x,\ldots,x^{n-1}]{\rm diag}(1,v^{n^\prime},\ldots, (v^{n^\prime})^{n-1})M_{a(x)}V$$
$(\forall a(x)\in R[x]/\langle x^{p^kn}-(1+\omega u)\rangle)$ is an isomorphism
of rings from $R[x]/\langle x^{p^kn}-(1+\omega u)\rangle$ onto $\mathcal{R}_k[x]/\langle x^n-1\rangle$. Therefore,
$C$ is a $(1+\omega u)$-constacyclic code of length $p^kn$ over $R$ if and only if
$\psi(\varphi(C))$ is a cyclic code of length $n$ over $\mathcal{R}_k$}.

\vskip 3mm \par
  Then by Lemma 3.4 and Theorem 2.2, we give a matrix-product structure of any $(1+\omega u)$-constacyclic code of length $p^kn$ over $R$ as follows.

\vskip 3mm \noindent
  {\bf Theorem 3.5} \textit{Using the notations above, let $C$ be a $(1+\omega u)$-constacyclic code of length $p^kn$ over $R$, assume $\mathcal{C}=\psi(\varphi(C))\subseteq \mathcal{R}_k[x]/\langle x^n-1\rangle$ and $C_s={\rm Tor}_s(\mathcal{C})\subseteq \mathbb{F}_{p^m}[x]/\langle x^n-1\rangle$ for all $s=0,1,\ldots,p^{k}e-1$. Denote}
$$\mathcal{C}_\rho=\bigoplus_{i=0}^{e-1}u^iC_{ip^k+\rho}\subseteq R[x]/\langle x^n-1\rangle, \ \rho=0,1,\ldots,p^k-1.$$

\par
  (i) \textit{$\mathcal{C}_\rho$ is a cyclic code of length $n$ over $R$ satisfying $|\mathcal{C}_\rho|=\prod_{i=0}^{e-1}|C_{ip^k+\rho}|$
for all $\rho=0,1,\ldots,p^k-1$. Moreover, we have that $\mathcal{C}_{p^k-1}\supseteq\ldots\supseteq \mathcal{C}_1\supseteq \mathcal{C}_0$}.

\par
  (ii) \textit{$\Theta(C)=M_{p^k}(n,\omega)\cdot C=[\mathcal{C}_{p^k-1},\mathcal{C}_{p^k-2}, \ldots,\mathcal{C}_1,\mathcal{C}_0]\cdot A_{p^k}$, where}
$$A_{p^k}=\left[(-1)^{p^k-i-j+1}\left(\begin{array}{c}p^k-i\cr j-1\end{array}\right)\right]_{1\leq i,j\leq p^k} \
({\rm mod} \ p)$$
\textit{in which we set $\left(\begin{array}{c}p^k-i\cr j-1\end{array}\right)=0$ if $p^k-i<j-1$ for all $1\leq i,j\leq p^k$}. \textit{Hence $C$ is monomially equivalent to
$[\mathcal{C}_{p^k-1},\mathcal{C}_{p^k-2},$ $\ldots,\mathcal{C}_1,\mathcal{C}_0]\cdot A_{p^k}$}.

\vskip 3mm \noindent
  \textit{Proof} (i) By Theorem 2.2, $C_s$ is a cyclic code of length $n$ over $\mathbb{F}_{p^m}$, $0\leq s\leq p^ke-1$, and satisfies
\begin{eqnarray*}
&&C_0\subseteq C_1\subseteq\ldots\subseteq C_{p^k-1}\subseteq C_{p^k}\subseteq C_{p^k+1}\subseteq\ldots\subseteq C_{2p^k-1}\\
&&\subseteq \ldots \subseteq C_{(e-1)p^k}\subseteq C_{(e-1)p^k+1}\subseteq\ldots\subseteq C_{ep^k-1}.
\end{eqnarray*}
This implies $C_{\rho}\subseteq C_{p^k+\rho}\subseteq \ldots \subseteq C_{(e-1)p^k+\rho}$. From this and by Corollary 2.3(ii),
we deduce that $\mathcal{C}_\rho=\bigoplus_{i=0}^{e-1}u^iC_{ip^k+\rho}$ is a cyclic code of length $n$ over $R$, i.e. an ideal of the ring $R[x]/\langle x^n-1\rangle$, satisfying $|\mathcal{C}_\rho|=\prod_{i=0}^{e-1}|C_{ip^k+\rho}|$.

\par
  Let $0\leq\rho<\rho^\prime\leq e-1$. Then $C_{ip^k+\rho}\subseteq C_{ip^k+\rho^\prime}$ for all $i=0,1,\ldots,e-1$. From this and by
Corollary 2.3(iii), we deduce
that $\mathcal{C}_\rho\subseteq\mathcal{C}_{\rho^\prime}$.

\par
  (ii) As $\omega\in R^\times$, for each integer $i$, $0\leq i\leq e-1$, $(\omega u)^i=\omega^iu^i$ can be uniquely expressed as
$(\omega u)^i=\sum_{j=i}^{e-1}\lambda_{i,j}u^j$ for some $\lambda_{i,j}\in \mathbb{F}_{p^m}$ where $\lambda_{i,i}\neq 0$, $\forall j=i,i+1,\ldots,e-1.$
Then we can write
\begin{equation}
\left[\begin{array}{c}1\cr \omega u \cr (\omega u)^2 \cr \ldots \cr (\omega u)^{e-1}\end{array}\right]
=TU \ {\rm with} \ U=[1,u,u^2,\ldots, u^{e-1}]^{{\rm tr}}=\left[\begin{array}{c}1\cr  u \cr u^2 \cr \ldots \cr u^{e-1}\end{array}\right],
\end{equation}
where $T=\left[\begin{array}{ccccc}\lambda_{0,0} & \lambda_{0,1} & \ldots & \lambda_{0,e-2}& \lambda_{0,e-1} \cr
 0 & \lambda_{1,1} & \ldots & \lambda_{1,e-2} & \lambda_{1,e-1} \cr
 \ldots & \ldots & \ldots &\ldots & \ldots \cr
 0 & 0 & \ldots & \lambda_{e-2,e-2}& \lambda_{e-2,e-1}\cr
  0 & 0 & \ldots & 0 & \lambda_{e-1,e-1}\end{array}\right]$
being an invertible $e\times e$ matrix over $\mathbb{F}_{p^m}$ (and $R$).
  By Theorem 2.2, we know that
$$\mathcal{C}=\psi\varphi(C)=C_0\oplus (v-1)C_1\oplus (v-1)^2C_2\oplus\ldots \oplus (v-1)^{p^ke-1}C_{p^ke-1}.$$

\par
  Let $a(x)=\sum_{j=0}^{n-1}\sum_{t=0}^{p^{k}-1}a_{j+tn}x^{j+tn}\in C$ where $a_{j+tn}\in R$, and assume $c(x)=\psi\varphi(a(x))\in \mathcal{C}$.
Then
  for each integer $s$, $0\leq s\leq p^{k}e-1$, there exists a unique codeword $c_s(x)\in C_s$ such that
$c(x)=\sum_{s=0}^{p^ke-1}(v-1)^sc_s(x)$.
By $(v-1)^{p^k}=\omega u$, we have $(v-1)^{ip^k}=(\omega u)^i$, for all $0\leq i\leq e-1$. Hence
\begin{equation}
c(x)=\sum_{i=0}^{e-1}\sum_{\rho=0}^{p^k-1}(v-1)^{i p^k+\rho}c_{ip^k+\rho}(x)
  =\sum_{\rho=0}^{p^k-1}(v-1)^\rho\sum_{i=0}^{e-1}(\omega u)^ic_{ip^k+\rho}(x)
\end{equation}
in which
\begin{eqnarray*}
\sum_{i=0}^{e-1}(\omega u)^ic_{ip^k+\rho}(x)&=&[c_{\rho}(x),c_{p^k+\rho}(x),\ldots,c_{p^k(e-1)+\rho}(x)]\left[\begin{array}{c}1\cr \omega u \cr (\omega u)^2 \cr \ldots \cr (\omega u)^{e-1}\end{array}\right]\\
 &=&[c_{\rho}(x),c_{p^k+\rho}(x),\ldots,c_{p^k(e-1)+\rho}(x)]TU
\end{eqnarray*}
by Equation (7). Let $0\leq \rho\leq p^k-1$. We denote the cartesian product of the $e$ cyclic codes $C_\rho, C_{p^k+\rho},\ldots, C_{p^k(e-1)+\rho}$ with length $n$ over $\mathbb{F}_{p^m}$ by $\mathcal{S}_\rho$, i.e.
$$\mathcal{S}_\rho=C_{\rho}\times C_{p^k+\rho}\times C_{2p^k+\rho}\times\ldots \times C_{p^k(e-1)+\rho}.$$
For any $[c_{\rho}(x),c_{p^k+\rho}(x),\ldots,c_{p^k(e-1)+\rho}(x)]\in \mathcal{S}_\rho$, we denote
$$[c^\prime_{\rho}(x),c^\prime_{p^k+\rho}(x),\ldots,c^\prime_{p^k(e-1)+\rho}(x)]
=[c_{\rho}(x),c_{p^k+\rho}(x),\ldots,c_{p^k(e-1)+\rho}(x)]T$$
where
$$c^\prime_{jp^k+\rho}(x)=\lambda_{0,j}c_{\rho}(x)+\lambda_{1,j}c_{p^k+\rho}(x)+\lambda_{2,j}c_{2p^k+\rho}(x)+\ldots+\lambda_{j,j}c_{jp^k+\rho}(x)$$
for all $j=0,1,\ldots,e-1$. By $C_{\rho}\subseteq C_{p^k+\rho}\subseteq C_{2p^k+\rho}\subseteq\ldots \subseteq C_{jp^k+\rho}$, it follows that
$$c^\prime_{jp^k+\rho}(x)\in C_{jp^k+\rho}, \ \forall j, \ 0\leq j\leq e-1,$$
and hence
$[c^\prime_{\rho}(x),c^\prime_{p^k+\rho}(x),\ldots,c^\prime_{p^k(e-1)+\rho}(x)]\in \mathcal{S}_\rho$. From this and by the invertibility of the matrix $T$, we deduce that
the map defined by
$$\xi\mapsto \xi\cdot T \ \ (\forall \xi=[c_{\rho}(x),c_{p^k+\rho}(x),\ldots,c_{p^k(e-1)+\rho}(x)]\in \mathcal{S}_\rho)$$
is a bijection on $\mathcal{S}_\rho$. This implies $\mathcal{S}_\rho=\{\xi\cdot T\mid \xi\in \mathcal{S}_\rho\}$.
Using the notations above, we have
\begin{eqnarray*}
 &&\sum_{i=0}^{e-1}(\omega u)^ic_{ip^k+\rho}(x)\\
  &=&[c^\prime_{\rho}(x),c^\prime_{p^k+\rho}(x),c^\prime_{2p^k+\rho}(x),\ldots,c^\prime_{p^k(e-1)+\rho}(x)]U\\
  &=&c^\prime_{\rho}(x)+uc^\prime_{p^k+\rho}(x)
  +u^2c^\prime_{2p^k+\rho}(x)+\ldots+u^{e-1}c^\prime_{(e-1)p^k+\rho}(x)\in \mathcal{C}_\rho,
\end{eqnarray*}
where
$$\mathcal{C}_\rho=C_{\rho}\oplus u C_{p^k+\rho}\oplus u^2 C_{2p^k+\rho}\oplus\ldots \oplus u^{e-1} C_{p^k(e-1)+\rho}\subseteq R[x]/\langle x^n-1\rangle.$$
Now, denote $\xi_\rho(x)=\sum_{i=0}^{e-1}(\omega u)^ic_{ip^k+\rho}(x)\in \mathcal{C}_\rho$ for all $\rho=0,1,\ldots,p^k-1$.
Then $c(x)=\sum_{\rho=0}^{p^k-1}(v-1)^\rho\xi_\rho(x)$. From this, by Equation (8) and
$$(v-1)^{p^k-i}=\left((-1)+v\right)^{p^k-i}=\sum_{j=1}^{p^k-i+1}\left(\begin{array}{c}p^k-i\cr j-1\end{array}\right)(-1)^{p^k-i-j+1}v^{j-1}$$
for all $i=1,2,\ldots, p^k$, we deduce that
\begin{equation}
c(x)=[1,x,\ldots,x^{n-1}][\xi_{p^{k}-1},\ldots,\xi_1,\xi_0]A_{p^k}V,
\end{equation}
where $\xi_\rho$ is the unique $n\times 1$ column vector over $R$ satisfying
$$\xi_\rho(x)=[1,x,\ldots,x^{n-1}]\cdot \xi_\rho, \ 0\leq \rho\leq p^k-1,$$
and $V=[1,v,v^2,\ldots,v^{p^k-1}]^{{\rm tr}}$.
From now on, we will identify $\xi_\rho(x)$ with $\xi_\rho$ as a codeword in
the cyclic code $\mathcal{C}_\rho$ over $R$ of length $n$.
   By replacing $v$ with $x^n$ in Equation (9) we obtain
\begin{equation}
\pi(c(x))=[1,x,\ldots,x^{n-1}][\xi_{p^{k}-1},\ldots,\xi_1,\xi_0]A_{p^k}X
\end{equation}
where $X=[1,x^n,\ldots,x^{(p^k-1)n}]^{{\rm tr}}$.

\par
   On the other hand, by Lemma 3.4 we have
$$c(x)=\psi\varphi(a(x))=[1,x,\ldots,x^{n-1}]{\rm diag}(1,v^{n^\prime},\ldots, (v^{n^\prime})^{n-1})M_{a(x)}V.$$
Replacing $v$ with $x^n$, we obtain
\begin{eqnarray*}
\pi(c(x))&=&[1,x,\ldots,x^{n-1}]{\rm diag}(1,(x^n)^{n^\prime},(x^n)^{2n^\prime},\ldots, (x^n)^{(n-1)n^\prime})M_{a(x)}X\\
 &=&[1,x^{1+n^\prime n},x^{2(1+n^\prime n)},\ldots,x^{(n-1)(1+n^\prime n)}]M_{a(x)}X\\
 &=&\sum_{j=0}^{n-1}\sum_{t=0}^{p^k-1}a_{j+tn}x^{j(1+n^\prime n)+tn}\\
 &=&\sum_{j=0}^{n-1}\sum_{t=0}^{p^k-1}a_{j+tn}x^{j+(t+jn^\prime)n}.
\end{eqnarray*}
By Notation 3.1, we have $n^\prime=qp^k+n^{\prime\prime}$, where $0\leq q\leq p^l-1$
and $1\leq n^{\prime\prime}\leq p^k-1$.
By $x^{p^kn}=1+\omega u$ in the ring $R[x]/\langle x^{p^kn}-(1+\omega u)\rangle$ it follows that
$$
a_{j+tn}x^{j+(t+jn^\prime)n}=x^{jq\cdot p^kn}a_{j+tn}x^{j+(t+jn^{\prime\prime})n}=(1+\omega u)^{jq} a_{j+tn}x^{j+(t+jn^{\prime\prime})n}.
$$
We denote $\lambda=t+jn^{\prime\prime}$ (mod $p^k$). Then $\lambda\in [p^k)$ and
$t=\lambda-jn^{\prime\prime}$  (mod $p^k$). By Notation 3.1 and Lemma 3.2,  the map
$\varrho$ defined by
$\varrho(j+\lambda n)=j+tn=j+n\left(\lambda-jn^{\prime\prime} \ ({\rm mod} \ p^k)\right)$ ($\forall
(j,\lambda)\in [n)\times [p^k)$)
is a permutation on the set $[p^kn)$. This implies
\begin{equation}
\pi(c(x))=\sum_{j=0}^{n-1}\sum_{\lambda=0}^{p^k-1}(1+\omega u)^{jq}a_{\varrho(j+\lambda n)}x^{j+\lambda n}
=[1,x,\ldots,x^{n-1}]\Lambda\widetilde{M}_{a(x)}X
\end{equation}
where $\Lambda={\rm diag}[1,(1+\omega u)^q,(1+\omega u)^{2q},\ldots,(1+\omega u)^{(n-1)q}]$ and
$$\widetilde{M}_{a(x)}=[b_{j,\lambda}] \
{\rm with} \ b_{j,\lambda}=a_{\varrho(j+\lambda n)}, \ \forall (j,\lambda)\in [n)\times[p^k).$$
Now, from Equations (10) and (11) we deduces that
$$\Lambda\cdot\widetilde{M}_{a(x)}=[\xi_{p^{k}-1},\ldots,\xi_1,\xi_0]A_{p^k}, \ \forall a(x)\in C.$$
In this paper, we regard $\Lambda\cdot\widetilde{M}_{a(x)}$ and $[\xi_{p^{k}-1},\ldots,\xi_1,\xi_0]A_{p^k}$ as a column vector of dimension $p^kn$ by reading the entries of
the matrix in column-major order respectively. According to this view, by Lemma 3.2 it follows
that
$$M_{p^k}(n,\omega)\cdot [a_0,a_1,\ldots,a_{p^kn-1}]^{{\rm tr}}
=[\xi_{p^{k}-1},\ldots,\xi_1,\xi_0]A_{p^k}, \forall a(x)\in C.$$

\par
 As stated above, we conclude that $\Theta(C)=[\mathcal{C}_{p^k-1},\mathcal{C}_{p^k-2}, \ldots,\mathcal{C}_1,\mathcal{C}_0]\cdot A_{p^k}$.
By Lemma 3.2(iii), $C$ and the matrix-product code $[\mathcal{C}_{p^{k}-1},\ldots,\mathcal{C}_1,\mathcal{C}_0]\cdot A_{p^k}$  are
monomially equivalent codes over the finite chain ring $R$.
\hfill
$\Box$

\vskip 3mm\par
  From Lemma 3.2 and the proof of Theorem 3.5, we deduce the following corollary
which will be used in the next section.

\vskip 3mm\noindent
  {\bf Corollary 3.6} \textit{Let $C\subseteq R^{p^kn}$. Then $C$ is a $(1+\omega u)$-constacyclic code
of length $p^kn$ over $R$ if and only if there is a sequence of cyclic codes $\mathcal{C}_{p^{k}-1}\supseteq\ldots\supseteq\mathcal{C}_1\supseteq\mathcal{C}_0$
over $R$ of length $n$ such that
$$M_{p^k}(n,\omega)\cdot C:=\{M_{p^k}(n,\omega)\cdot c\mid c\in C\}=[\mathcal{C}_{p^{k}-1},\ldots,\mathcal{C}_1,\mathcal{C}_0]\cdot A_{p^k},$$
where we regard each $c\in C$ as a $p^kn\times 1$ column vector over $R$ and each codeword
$\xi=[\xi_{p^k-1},\ldots,\xi_1,\xi_0]A_{p^k}$ in $[\mathcal{C}_{p^{k}-1},\ldots,\mathcal{C}_1,\mathcal{C}_0]\cdot A_{p^k}$ as a $p^kn\times 1$ column vector over $R$
by reading the entries of the matrix $\xi$ in column-major order, respectively.}

\vskip 3mm \par
     As ${\rm gcd}(p,n)=1$, there are pairwise coprime monic irreducible polynomials $f_1(x),f_2(x),\ldots,f_r(x)$ in $\mathbb{F}_{p^m}[x]$
such that $x^n-1=f_1(x)f_2(x)\ldots f_r(x).$

\vskip 3mm \noindent
  {\bf Lemma 3.7} (cf. [4] Theorem 3.4) \textit{Using the notations above, all $(p^{k}e+1)^r$ distinct $(1+\omega u)$-constacyclic codes of length $p^kn$ over $R$ are given by}
$$C_{(i_1,i_2,\ldots,i_r)}=\langle f_1(x)^{i_1}f_2(x)^{i_2}\ldots f_r(x)^{i_r}\rangle\subseteq R[x]/\langle x^{p^kn}-(1+\omega u)\rangle,$$
\textit{where $0\leq i_1, i_2,\ldots, i_r\leq p^{k}e$.
Furthermore, the number of codewords in $C_{(i_1,i_2,\ldots,i_r)}$ is equal to}
$|C_{(i_1,i_2,\ldots,i_r)}|=p^{m(\sum_{t=1}^r(p^{k}e-i_t){\rm deg}(f_t(x)))}$.

\vskip 3mm\par
   Finally, we determine the nested sequences of $p^{k}$
cyclic codes $\mathcal{C}_{p^{k}-1}\supseteq\ldots\supseteq \mathcal{C}_1\supseteq \mathcal{C}_0$ with length $n$ over $R$ in the matrix-product
structure of
a $(1+\omega u)$-constacyclic code $C_{(i_1,i_2,\ldots,i_r)}$ over $R$ with length $p^kn$.

\vskip 3mm \noindent
  {\bf Theorem 3.8} \textit{Using the notations above, let $C=\langle f_1(x)^{i_1}f_2(x)^{i_2}\ldots f_r(x)^{i_r}\rangle$, $0\leq i_1,i_2,\ldots,i_r\leq p^{k}e$, being a $(1+\omega u)$-constacyclic code $C$ of length $p^kn$ over $R$. For each integer $s$, $0\leq s\leq p^ke-1$, denote}
$$g_s(x)=\prod_{i_t>s, 1\leq t\leq r}f_t(x)\in \mathbb{F}_{p^m}[x].$$
 \textit{Then $C$ is monomially equivalent to $[\mathcal{C}_{p^{k}-1}, \ldots, \mathcal{C}_1,\mathcal{C}_0]\cdot A_{p^k}$, where $A_{p^k}$ is given by Theorem 3.5 and for each integer $\rho$, $0\leq \rho\leq p^k-1$, $\mathcal{C}_\rho$ is a cyclic code of length $n$ over
$R$ given by}
\begin{eqnarray*}
\mathcal{C}_\rho &=& \left\langle g_{\rho}(x),ug_{p^k+\rho}(x),u^2g_{2p^k+\rho}(x),\ldots,u^{e-1}g_{(e-1)p^k+\rho}(x)\right\rangle\\
  &=&\left\langle g_{\rho}(x)+ug_{p^k+\rho}(x)+u^2g_{2p^k+\rho}(x)+\ldots+u^{e-1}g_{(e-1)p^k+\rho}(x)\right\rangle.
\end{eqnarray*}

\vskip 3mm \noindent
  \textit{Proof} Denote $G(x)=f_1(x)^{i_1}f_2(x)^{i_2}\ldots f_r(x)^{i_r}\in \mathbb{F}_{p^m}[x]$. By Theorem 3.5, it is suffices to prove that $C_s=\langle g_s(x)\rangle$ for all $s=0,1,\ldots,p^ke-1$.
Let $0\leq s\leq p^ke-1$. We first verify that
$$g_s(x)\in C_s={\rm Tor}_s(\psi(\varphi(C)))=\tau(\psi(\varphi(C)):(v-1)^s),$$
which is equivalent to that
$(v-1)^s\left(g_s(x)+(v-1)w(x)\right)\in \psi(\varphi(C))$
for some $w(x)\in \mathcal{R}_k[x]/\langle x^n-1\rangle$.
  In the following, we denote
$$A_s=\{t\mid i_t>s, \ 1\leq t\leq r\}, \  B_s=\{t\mid i_t\leq s, \ 1\leq t\leq r\}.$$
and set
$$h_s(x)=\prod_{t\in B_s}f_t(x), \ \widehat{f}_s(x)=\prod_{t\in A_s}f_t(x)^{i_t-s-1}.$$
Then $g_s(x)=\prod_{t\in A_s}f_t(x)$, and $h_s(x), \widehat{f}_s(x)\in \mathbb{F}_{p^m}[x]$ satisfying $x^n-1=g_s(x)h_s(x)$,
${\rm gcd}(g_s(x),h_s(x))={\rm gcd}(\widehat{f}_s(x),h_s(x))=1$.

\par
   As $G(x)=\prod_{t\in A_s\cup B_s}f_t(x)^{i_t}$ and $i_t\leq s$ for all $t\in B_s$, we have
$$(x^n-1)^sg_s(x)\widehat{f}_s(x)=\prod_{t\in A_s\cup B_s}f_t(x)^s\prod_{t\in A_s}f_t(x)\prod_{t\in A_s}f_t(x)^{i_t-s-1}
=\varepsilon(x)G(x)$$
where $\varepsilon(x)=\prod_{t\in B_s}f_t(x)^{s-i_t}\in\mathbb{F}_{p^m}[x]\subseteq R[x]$. This implies
\begin{equation}
(x^n-1)^sg_s(x)\widehat{f}_s(x)\in \langle G(x)\rangle=C.
\end{equation}
By ${\rm gcd}(\widehat{f}_s(x),h_s(x))=1$, there exist $a(x),b(x)\in \mathbb{F}_{p^m}[x]$ such that
$a(x)\widehat{f}_s(x)+b(x)h_s(x)=1$. This implies $a(x)\widehat{f}_s(x)=1-b(x)h_s(x)$. Then by Equation (12) and $x^n-1=g_s(x)h_s(x)$, it follows
that
\begin{eqnarray*}
&&(x^n-1)^sg_s(x)-(x^n-1)^{s+1}b(x)\\
 &=&(x^n-1)^sg_s(x)-(x^n-1)^{s}\cdot g_s(x)h_s(x)\cdot b(x)\\
 &=&(x^n-1)^sg_s(x)(1-b(x)h_s(x))\\
 &=&(x^n-1)^sg_s(x)\widehat{f}_s(x)\cdot a(x)\in C.
\end{eqnarray*}
Replacing $x^n$ with $v$, by the definition of $\varphi$ we obtain
\begin{eqnarray*}
&&(v-1)^sg_s(x)-(v-1)^{s+1}\varphi(b(x))\\
 &=&\varphi\left((x^n-1)^sg_s(x)-(x^n-1)^{s+1}b(x)\right)\in \varphi(C).
\end{eqnarray*}
This implies $(v-1)^sg_s(x)+(v-1)^{s+1}\alpha(x)\in \varphi(C)$, where $\alpha(x)=-\varphi(b(x))\in
\mathcal{R}_k[x]/\langle x^n-v\rangle$. Then we replace $x$ with $v^{n^\prime}x$, by the definition
of $\psi$ we have
\begin{eqnarray*}
&&(v-1)^sg_s(v^{n^\prime}x)+(v-1)^{s+1}\alpha(v^{n^\prime}x)\\
&=&\psi\left((v-1)^sg_s(x)+(v-1)^{s+1}\alpha(x)\right)\in \psi(\varphi(C)).
\end{eqnarray*}
From this and by
$$g_s(v^{n^\prime}x)=g_s(x+x(v^{n^\prime}-1))=g_s(x+(v-1)\beta(x))=g_s(x)+(v-1)\delta(x)$$
for some $\delta(x)\in \mathcal{R}_k[x]/\langle x^n-1\rangle$, where $\beta(x)=x\sum_{i=0}^{n^\prime-1}v^i$, we deduce that
\begin{eqnarray*}
&&(v-1)^sg_s(v^{n^\prime}x)+(v-1)^{s+1}\alpha(v^{n^\prime}x)\\
&=&(v-1)^s\left(g_s(x)+(v-1)\delta(x)\right)+(v-1)^{s+1}\alpha(v^{n^\prime}x)\\
&=&(v-1)^s\left(g_s(x)+(v-1)(\delta(x)+\alpha(v^{n^\prime}x))\right).
\end{eqnarray*}
This implies $g_s(x)+(v-1)(\delta(x)+\alpha(v^{n^\prime}x))\in (\psi(\varphi(C)):(v-1)^s)$, and hence
$$g_s(x)=\tau\left(g_s(x)+(v-1)(\delta(x)+\alpha(v^{n^\prime}x))\right)\in \tau(\psi(\varphi(C)):(v-1)^s)
=C_s.$$
Therefore, $\langle g_s(x)\rangle\subseteq C_s$ as ideals of the ring $\mathbb{F}_{p^m}[x]/\langle x^n-1\rangle$ for all $s=0,1,\ldots,p^ke-1$.

\par
  On the other hand, by $x^n-1=\prod_{t=1}^rf_t(x)$ and $G(x)=\prod_{t=1}^rf_t(x)^{i_t}=\prod_{s=0}^{p^{k}e-1}g_s(x)$
it follows that
\begin{equation}
\sum_{t=1}^r(p^{k}e-i_t){\rm deg}(f_t(x))=p^{k}en-{\rm deg}(G(x))=\sum_{s=0}^{p^{k}e-1}(n-{\rm deg}(g_s(x))).
\end{equation}
By Lemma 3.7, Theorem 2.2 and $|\langle g_s(x)\rangle|=p^{m(n-{\rm deg}(g_s(x)))}$ for all $s$, we have
\begin{eqnarray*}
p^{m(\sum_{t=1}^r(p^{k}e-i_t){\rm deg}(f_t(x)))}&=&
|C|=|\psi(\varphi(C))|=\prod_{s=0}^{p^ke-1}|C_s|\\
  &\geq&\prod_{s=0}^{p^ke-1}|\langle g_s(x)\rangle|=\prod_{s=0}^{p^ke-1}(p^m)^{n-{\rm deg}(g_s(x))}.
\end{eqnarray*}
From this and by Equation (13), we deduce that $C_s=\langle g_s(x)\rangle$,
i.e. $g_s(x)$ is the generator polynomial of $C_s$ for all $s$.

\par
   Finally, let $0\leq \rho\leq p^k-1$. By Corollary 2.3(ii),
it follows that $\mathcal{C}_\rho=\bigoplus_{i=0}^{e-1}u^iC_{ip^k+\rho}=\langle g_\rho(x)+ug_{p^k+\rho}(x)+u^2g_{2p^k+\rho}(x)+\ldots+u^{e-1}g_{p^k(e-1)+\rho}(x)\rangle$
as ideals of $R[x]/\langle x^n-1\rangle$.
\hfill
$\Box$

\vskip 3mm\noindent
  \textit{Remark} As $\mathcal{C}_{p^{k}-1}\supseteq\ldots\supseteq\mathcal{C}_1\supseteq\mathcal{C}_0$,
by Theorem 2.1 the minimum Hamming distance of the $(1+\omega u)$-constacyclic code $C$ of length $p^kn$ over $R$ is equal to
$d={\rm min}\{\delta_{i+1}d_i\mid i=0,1,\ldots,p^k-1\}$, where
$d_i$ is the minimum Hamming distance of the cyclic code $\mathcal{C}_i$ of length $n$ over $R$ and
 $\delta_{i+1}$ is the minimum distance of the linear code $\mathcal{L}_{i+1}$ with length $p^k$ over $\mathbb{F}_{p^m}$ generated by the first $i+1$ rows of
the matrix $A_{p^k}$, for all $i=0,1,\ldots,p^k-1$. For each integer $1\leq j\leq p^k$, it can be easily seen that $\mathcal{L}_j$ is exactly the cyclic code of length $p^k$ over $\mathbb{F}_{p^m}$ generated by $(x-1)^{p^k-j}$. Since $a(x)\mapsto a(-x)$
($\forall a(x)\in \mathbb{F}_{p^m}[x]/\langle x^{p^k}-1\rangle$) is a ring isomorphism and a Hamming
distance-preserving map from $\mathbb{F}_{p^m}[x]/\langle x^{p^k}-1\rangle$ onto $\mathbb{F}_{p^m}[x]/\langle x^{p^k}+1\rangle$, $\delta_j$ is equal to the minimum Hamming
distance of the negacyclic code of length $p^k$ over $\mathbb{F}_{p^m}$ generated by $(x+1)^{p^k-j}$.
From this and by Dinh [9] Theorem 4.11, we deduce that
$$
\delta_j=\left\{\begin{array}{ll}
(t+1)p^s, & {\rm if} \ p^{k-s}-tp^{k-s-1}\leq j\leq p^{k-s}-tp^{k-s-1}+p^{k-s-1}-1,\cr
          & {\rm where} \ 1\leq t\leq p-1 \ {\rm and} \ 1\leq s\leq k-1; \cr
\gamma+1,  & {\rm if} \ p^k-\gamma p^{k-1}\leq j\leq p^k-\gamma p^{k-1}+p^{k-1}-1,\cr
         & {\rm where} \ 1\leq \gamma\leq p-1; \cr
1,        & {\rm if} \ j=p^k.
\end{array}\right.$$



\section{Iterative construction of $(1+\omega u)$-constacyclic codes over $R$}
\noindent
   Let $A=[a_{ij}]_{1\leq i,j\leq m}$ and $B=[b_{st}]_{1\leq s,t\leq n}$ be $m\times m$ and $n\times n$
matrices over a commutative ring $\Gamma$, respectively. The \textit{Kronecker product} of $A$ and $B$ is defined by
$A\otimes B=[a_{ij}B]_{1\leq i,j\leq m}=\left[\begin{array}{cccc}a_{11}B & a_{12}B & \ldots & a_{1m}B\cr
 a_{21}B & a_{22}B & \ldots & a_{2m}B \cr \ldots &\ldots &\ldots &\ldots \cr a_{m1}B & a_{m2}B & \ldots & a_{mm}B\end{array}\right]$
where
$a_{ij}B=\left[\begin{array}{cccc}a_{ij}b_{11} & a_{ij}b_{12} & \ldots & a_{ij}b_{1n}\cr
 a_{ij}b_{21} & a_{ij}b_{22} & \ldots & a_{ij}b_{2n} \cr \ldots &\ldots &\ldots &\ldots \cr a_{ij}b_{n1} & a_{ij}b_{n2} & \ldots & a_{ij}b_{nn}\end{array}\right],
\ i,j=1,\ldots,m.$
It is known from linear algebra that if $A, B, C, D$ are matrix of appropriate sizes, then
$(A\otimes B)\otimes C=A\otimes(B\otimes C)$ and $(A\otimes B)(C\otimes D)=(AC)\otimes(BD)$.

\par
   Using the notation in Theorem 3.5, we know that
$$A_{p^k}=\left[(-1)^{p^k-i-j+1}\left(\begin{array}{c}p^k-i\cr j-1\end{array}\right)\right]_{1\leq i,j\leq p^k} \
({\rm mod} \ p), \ \forall k\geq 1.$$
Especially, $A_p=\left[(-1)^{p-i-j+1}\left(\begin{array}{c}p-i\cr j-1\end{array}\right)\right]_{1\leq i,j\leq p}$
(mod $p$), i.e.
$$A_p=\left[\begin{array}{cccccc}(-1)^{p-1} & (-1)^{p-2}\left(\begin{array}{c}p-1\cr p-2\end{array}\right) &
\ldots & (-1)^2\left(\begin{array}{c}p-1\cr 2\end{array}\right) & -\left(\begin{array}{c}p-1 \cr 1\end{array}\right) & 1 \cr
(-1)^{p-2} & (-1)^{p-3}\left(\begin{array}{c}p-2\cr p-3\end{array}\right) &
\ldots & -\left(\begin{array}{c}p-2\cr 1\end{array}\right) & 1 & 0 \cr
(-1)^{p-3} & (-1)^{p-4}\left(\begin{array}{c}p-3\cr p-4\end{array}\right) &
\ldots & 1 & 0 & 0 \cr
-1 & 1 & \ldots  & 0 & 0 & 0 \cr
1 & 0 & \ldots  & 0 & 0 & 0
\end{array}\right]$$
(mod $p$). As $p$ is prime, by induction on $j$ it follows that $\left(\begin{array}{c}p-i\cr j-1\end{array}\right)\not\equiv 0$
(mod $p$), i.e. $\left(\begin{array}{c}p-i\cr j-1\end{array}\right)\in \mathbb{F}_{p^m}^\times$, for all integers $i, j$ satisfying $i+j\leq p+1$.
Moreover, by [17] Proposition 1 and Lemma 3 we know the following conclusions.

\vskip 3mm\noindent
  {\bf Lemma 4.1} (i) \textit{The matrix $A_p$ is an NSC matrix over $\mathbb{F}_{p^m}$}.

\par
  (ii) \textit{$A_{p^k}=A_p\otimes A_{p^{k-1}}=\left[(-1)^{p-i-j+1}\left(\begin{array}{c}p-i\cr j-1\end{array}\right) A_{p^{k-1}}\right]_{1\leq i,j\leq p}$ $({\rm mod} \ p)$ for any integer $k\geq 2$}.

\vskip 3mm \par
   Every $(1+\omega u)$-constacyclic code of length
$p^kn$ over $R=\mathbb{F}_{p^m}[u]/\langle u^e\rangle$ can be constructed recursively from
$(1+\omega u)$-constacyclic codes of length
$p^{k-1}n$ over $R$ by the following theorem.

\vskip 3mm \noindent
  {\bf Theorem 4.2} \textit{Let $C$ be a $(1+\omega u)$-constacyclic code
of length $p^kn$ over $R$. Then there exist $(1+\omega u)$-constacyclic codes $C^{(p-1)},\ldots,C^{(1)}, C^{(0)}$
of length $p^{k-1}n$ over $R$ satisfying $C^{(p-1)}\supseteq \ldots\supseteq C^{(1)}\supseteq C^{(0)}$ such that
$C$ is monomially equivalent to the following matrix-product code over $R$}
$$[C^{(p-1)},\ldots,C^{(1)}, C^{(0)}]\cdot A_p.$$
  \textit{Specifically, if $C=\langle f_1(x)^{i_1}f_2(x)^{i_2}\ldots f_r(x)^{i_r}\rangle$
as an ideal of $R[x]/\langle x^{p^kn}-(1+\omega u)\rangle$, where $0\leq i_1, i_2,\ldots,i_r\leq p^ke$, then}
\begin{equation}
C^{(j)}=\left\langle \prod_{\lambda=0}^{e-1}\prod_{l=0}^{p^{k-1}-1}
\prod_{i_t> \lambda p^k+jp^{k-1}+l, \ 1\leq t\leq r}f_{t}(x)\right\rangle_{R[x]/\langle x^{p^{k-1}n}-(1+\omega u)\rangle}
\end{equation}
\textit{for all $j=0,1,\ldots,p-1$. Furthermore, the minimum Hamming distance of $C$ is equal to}
${\rm min}\{pd^{(p-1)}, (p-1)d^{(p-2)},\ldots,2d^{(1)},d^{(0)}\}$
\textit{where $d^{(j)}$ is the minimum Hamming distance of $C^{(j)}$ for all $j=0,1,\ldots,p-1$.}

\vskip 3mm \noindent
  \textit{Proof} Let $C$ be a $(1+\omega u)$-constacyclic code of length $p^kn$ over $R$. By Theorem 3.5
and Lemma 4.1(ii),
there are $p^k$ cyclic codes $\mathcal{D}_{p^k-1},\ldots,\mathcal{D}_1,\mathcal{D}_0$ of length $n$ over $R$ satisfying
$\mathcal{D}_{p^k-1}\supseteq\ldots\supseteq\mathcal{D}_1\supseteq\mathcal{D}_0$
such that $C$ is monomially equivalent to the following matrix-product code
\begin{eqnarray*}
&&[\mathcal{D}_{p^k-1},\ldots,\mathcal{D}_1,\mathcal{D}_0]\cdot A_{p^k}\\
&=&\left[\mathcal{D}_{(p-1)p^{k-1}+p^{k-1}-1},
\ldots,\mathcal{D}_{(p-1)p^{k-1}+1},\mathcal{D}_{(p-1)p^{k-1}},\ldots,\mathcal{D}_{p^{k-1}-1}, \right.\\
&&\left. \ldots,\mathcal{D}_1,\mathcal{D}_0\right]\cdot \left[(-1)^{p-1-s-t}\left(\begin{array}{c}p-s\cr t-1\end{array}\right)A_{p^{k-1}}\right]_{1\leq s,t\leq p} \ ({\rm mod} \ p) \\
&=&[\mathcal{D}^{(p-1)},\ldots,\mathcal{D}^{(1)}, \mathcal{D}^{(0)}]\cdot A_p,
\end{eqnarray*}
where
$\mathcal{D}^{(j)}=[\mathcal{D}_{jp^{k-1}+p^{k-1}-1},\mathcal{D}_{jp^{k-1}+p^{k-1}-2},\ldots,\mathcal{D}_{jp^{k-1}+1},
\mathcal{D}_{jp^{k-1}}]\cdot A_{p^{k-1}}$ for all $j=0,1,\ldots,p-1$.
By Theorem 3.5, $\mathcal{D}^{(j)}$ is monomially equivalent to a $(1+\omega u)$-constacyclic code $C^{(j)}$ of length $p^{k-1}n$ over $R$
for all $j=0,1,\ldots,p-1$.
  Then by Corollary 3.6 and Notation 3.1, there is a
fixed $p^{k-1}n\times p^{k-1}n$ monomial matrix $M_{p^{k-1}}(n,\omega)$ over $R$ such that $M_{p^{k-1}}(n,\omega)\cdot C^{(j)}=\mathcal{D}^{(j)}$ for all $j$. This implies
\begin{eqnarray*}
&&[\mathcal{D}^{(p-1)},\ldots,\mathcal{D}^{(1)}, \mathcal{D}^{(0)}]\cdot A_p\\
&=&[M_{p^{k-1}}(n,\omega)\cdot C^{(p-1)},\ldots,M_{p^{k-1}}(n,\omega)\cdot C^{(1)}, M_{p^{k-1}}(n,\omega)\cdot C^{(0)}]\cdot A_p\\
&=&\left(I_p\otimes M_{p^{k-1}}(n,\omega)\right)\cdot\left([C^{(p-1)},\ldots,C^{(1)}, C^{(0)}]\cdot A_p\right)
\end{eqnarray*}
in which we regard $[C^{(p-1)},\ldots,C^{(1)}, C^{(0)}]\cdot A_p$ as a $p^kn\times 1$ column vector over $R$
by reading the entries of the matrix in column-major order, and $I_p$ is the identity matrix of order $p$.
Since
$$I_p\otimes M_{p^{k-1}}(n,\omega)={\rm diag}(M_{p^{k-1}}(n,\omega),\ldots, M_{p^{k-1}}(n,\omega))$$
 is a $p^kn\times p^kn$ monomial matrix over $R$, $[\mathcal{D}^{(p-1)},\ldots,\mathcal{D}^{(1)}, \mathcal{D}^{(0)}]\cdot A_p$
is monomially equivalent to $[C^{(p-1)},\ldots,C^{(1)}, C^{(0)}]\cdot A_p$. Moreover, by
$$\mathcal{D}_{jp^{k-1}+s}\supseteq \mathcal{D}_{ip^{k-1}+s}, \ \forall i,j,s, \ p-1\geq j>i\geq 0, \ s=0,1,\ldots, p^{k-1}-1$$
we
conclude that $\mathcal{D}^{(j)}\supseteq \mathcal{D}^{(i)}$. This implies  $C^{(j)}\supseteq C^{(i)}$ for all $0\leq i\leq j\leq p-1$.

\par
  Since $C=\langle f_1(x)^{i_1}f_2(x)^{i_2}\ldots f_r(x)^{i_r}\rangle$ which is an ideal
of $R[x]/\langle x^{p^kn}-(1+\omega u)\rangle$, in the matrix-product code
$[\mathcal{D}_{p^k-1},\ldots,\mathcal{D}_1,\mathcal{D}_0]\cdot A_{p^k}$ we have
$\mathcal{D}_\rho = \left\langle g_{\rho}(x),ug_{p^k+\rho}(x),u^2g_{2p^k+\rho}(x),\ldots,u^{e-1}g_{(e-1)p^k+\rho}(x)\right\rangle
\subseteq R[x]/\langle x^n-1\rangle$ for all $\rho=0,1,\ldots,p^k-1$. By Theorem 3.8, we see that $g_s(x)=\prod_{i_t>s, 1\leq t\leq r}f_t(x)\in \mathbb{F}_{p^m}[x]$,
$0\leq s\leq p^ke-1$, satisfying
$f_1(x)^{i_1}f_2(x)^{i_2}\ldots f_r(x)^{i_r}=\prod_{s=0}^{p^ke-1}g_s(x)$ $=\prod_{\eta=0}^{p^k-1}\prod_{\lambda=0}^{e-1}g_{\lambda p^k+\eta}(x).$
Let $0\leq j\leq p-1$.
Using Theorem 3.8 for the code $\mathcal{D}^{(j)}=[\mathcal{D}_{jp^{k-1}+p^{k-1}-1},\mathcal{D}_{jp^{k-1}+p^{k-1}-2},\ldots,\mathcal{D}_{jp^{k-1}+1},
\mathcal{D}_{jp^{k-1}}]\cdot A_{p^{k-1}}$, we deduce that $\mathcal{D}^{(j)}$ is monomially equivalent to the $(1+\omega u)$-constacyclic code $C^{(j)}$ of length $p^{k-1}n$ over $R$ generated by the following polynomial
$$G^{(j)}(x)=\prod_{l=0}^{p^{k-1}-1}\prod_{\lambda=0}^{e-1}g_{\lambda p^k+jp^{k-1}+l}(x)=\prod_{\lambda=0}^{e-1}\prod_{l=0}^{p^{k-1}-1}
\prod_{i_t> \lambda p^k+jp^{k-1}+l, \ 1\leq t\leq r}f_{t}(x).$$
Hence $C^{(j)}=\langle G^{(j)}(x)\rangle$ as ideals of $R[x]/\langle x^{p^{k-1}n}-(1+\omega u)\rangle$.

\par
  Finally, the conclusion for minimum Hamming distance of $C$ follows from Theorem 2.1 and Lemma 4.1(i) immediately.
\hfill
$\Box$

\vskip 3mm\noindent
   \textit{Remark} Let $k=0$. As ${\rm gcd}(p,n)=1$ and $(1+\omega u)^{p^l}=1$ by $p^l\geq e$ and $u^e=0$ in $R$,
there is uniquely $\eta\in R^\times$ such that $\eta^n=1+\omega u$. This implies
$(\eta x)^n=(1+\omega u) x^n$. Hence the map $\tau: a(x)\mapsto a(\eta x)$ ($\forall a(x)\in R[x]/\langle x^n-1\rangle$)
is an isomorphism of rings from $R[x]/\langle x^n-1\rangle$ onto $R[x]/\langle x^n-(1+\omega u)\rangle$. Therefore,
$C$ is a $(1+\omega u)$-constacyclic code of length $n$  over $R$ if and only if there is a unique cyclic code $D$ of length $n$ over $R$
such that $\tau(D)=C$. Obviously, $D$ and $\tau(D)$ are monomially equivalent codes over $R$.



\section{An Example}
\noindent
  In this section, we explain the main results of the paper by considering $(1+u)$-constacyclic codes of length $90$ over $R=\mathbb{F}_3+u\mathbb{F}_3$ ($u^2=0$).
In this case, we have $p=3$, $m=1$, $e=2$, $k=2$, $\omega=1\in R^\times$ and $n=10$.

\par
  Using Notation 3.1, by $3^1>e$ we have $l=1$, $p^{k+l}=3^{2+1}=27$,
and $n^\prime=19$ satisfying $1\leq n^\prime\leq 26$ and $n^\prime n=190\equiv 1$ (mod $27$).
Obviously,
$n^\prime=qp^k+n^{\prime\prime}$ where $p^k=9$, $q=2$ and $n^{\prime\prime}=1$. Hence the permutation
$\varrho$ on the set $[90)=\{0,1,\ldots,89\}$ is defined by
$$\varrho(j+10\lambda)=j+10(\lambda-j \ {\rm mod} \ 9),$$
for all $0\leq j\leq 9$ and $0\leq \lambda\leq 8$. Precisely, we have
\begin{eqnarray*}
\varrho&=&\left(\begin{array}{ccccc ccccc ccccc ccccc} 0 & 1 & 2 & 3 & 4 & 5 & 6 & 7 & 8 & 9 & 10 & 11 & 12 & 13 & 14 & 15 & 16 & 17 & 18 & 19 \cr
0  & 81 & 72 & 63 & 54 & 45 & 36 & 27 & 18 & 9  &
10 & 1  & 82 & 73 & 64 & 55 & 46 & 37 & 28 & 19
\end{array}\right. \\
&&\left.\begin{array}{ccccc ccccc c ccccc ccccc}
20 & 21 & 22 & 23 & 24 & 25 & 26 & 27 & 28 & 29 & \ldots & 80 & 81 & 82 & 83 & 83 & 85 & 86 & 87 & 88 & 89\cr
20 & 11 & 2 & 83 & 74 & 65 & 56 & 47 & 38 & 29 & \ldots &
80 &71  & 62 & 53 & 44 & 35 & 26 & 17 & 8  & 89\end{array}\right).
\end{eqnarray*}
By $(1+u)^2=1+2u$, it follows that $\Lambda={\rm diag}[1,1+2u, (1+2u)^2,\ldots, (1+2u)^9]$.
As $(1+u)^3=1$, we have
$$\Lambda
=\left[\begin{array}{cccc}
 1 & & & \cr
   & \Omega & & \cr
   & &  \Omega & \cr
   & & & \Omega
\end{array}\right] \
{\rm where} \ \Omega=\left[\begin{array}{ccc} 1+2u & & \cr & 1+u & \cr & & 1 \end{array}\right].$$
Let $P_{90}=[\epsilon_{i,j}]$ be the $90\times 90$ permutation matrix defined by:
$\epsilon_{i,j}=1$ if $j=\varrho(i)$, and $\epsilon_{i,j}=0$ othwise,
for all $0\leq i,j\leq 89$, and
set
$$M_9(10,1)={\rm diag}[\stackrel{9^{,}{\rm s}}{\overbrace{\Lambda,\ldots,\Lambda}}]\cdot P_{90}.$$
By Lemma 3.2, we see that $M_9(10,1)$ is a $90\times 90$ monomial matrix over $R$, and
$$\Theta(\left[\begin{array}{c} a_0 \cr a_1 \cr \ldots \cr a_{89}\end{array}\right])
=M_9(10,1)\left[\begin{array}{c} a_0 \cr a_1 \cr \ldots \cr a_{89}\end{array}\right],
\ \forall a_i\in R, \ i=0,1,\ldots, 89.$$
defines an $R$-module automorphism on $R^{90}$. By Corollary 3.6, we know that
$C$ is a $(1+u)$-constacyclic code of length $90$ over $R$ if and only if there is
a sequence $\mathcal{C}_8\supseteq\ldots\supseteq\mathcal{C}_1\supseteq\mathcal{C}_0$ of
cyclic codes of length $10$ over $R$ such that
$$
\Theta(C)=\{\Theta(c)\mid c\in C\}=[\mathcal{C}_8,\mathcal{C}_7,\mathcal{C}_6,\mathcal{C}_5,
  \mathcal{C}_5,\mathcal{C}_4,\mathcal{C}_3,\mathcal{C}_2,\mathcal{C}_1,\mathcal{C}_0]\cdot A_9.$$

\par
  Obviously, we have that $x^{10}-1=f_1(x)f_2(x)f_3(x)f_4(x)$ where
$f_1(x)=x+1$, $f_2(x)=x+2$, $f_3(x)=x^4+x^3+x^2+x+1$ and $f_4(x)=x^4+2x^3+x^2+2x+1$
being irreducible polynomials in $\mathbb{F}_3[x]$. By Lemma 3.7, the number of $(1+u)$-constacyclic codes with length $90$ over $R$
is $(3^2\cdot 2+1)^4=130321$ and all these codes are given by:
$$C_{(i_1,i_2,i_3,i_4)}=\left\langle f_1(x)^{i_1}f_2(x)^{i_2}f_3(x)^{i_3}f_4(x)^{i_4}\right\rangle_{R[x]/\langle x^{90}-(1+u)\rangle}$$
where $0\leq i_1,i_2,i_3,i_4\leq 18$. The number of codewords in $C_{(i_1,i_2,i_3,i_4)}$ is equal to
$|C_{(i_1,i_2,i_3,i_4)}|=3^{(18-i_1)+(18-i_2)+4(18-i_3)+4(18-i_4)}=3^{180-(i_1+i_2+4i_3+4i_4)}$.

\par
   Now, we consider $C=C_{(7,2,18,15)}=\langle f_1(x)^{7}f_2(x)^{2}f_3(x)^{18}f_4(x)^{15}\rangle$.
In this case, $(i_1,i_2,i_3,i_4)=(7,2,18,15)$, and hence $|C|=3^{180-(7+2+4\cdot 18+4\cdot 15)}=3^{39}$.

\par
   Using the notations of Theorem 3.8, we have
$g_s(x)=\prod_{i_t>s, 1\leq t\leq 4}f_t(x)\in \mathbb{F}_{3}[x]$ for all $s=0,1,\ldots,17$. Specifically, we have

\par
  $g_0(x)=g_1(x)=f_1(x)f_2(x)f_3(x)f_4(x)=x^{10}-1$,

\par
  $g_2(x)=g_3(x)=g_4(x)=g_5(x)=g_6(x)=f_1(x)f_3(x)f_4(x)=x^9+x^8+x^7+x^6+x^5+x^4+x^3+x^2+x+1$,

\par
  $g_7(x)=g_8(x)=g_9(x)=g_{10}(x)=g_{11}(x)=g_{12}(x)=g_{13}(x)=g_{14}(x)=f_3(x)f_4(x)=x^8+x^6+x^4+x^2+1$,

\par
  $g_{15}(x)=g_{16}(x)=g_{17}(x)=f_3(x)$.

\par
  $\bullet$ By Theorem 3.5, we have that $\Theta(C)=[\mathcal{C}_8,\mathcal{C}_7,\mathcal{C}_6,\mathcal{C}_5,\mathcal{C}_4,\mathcal{C}_3,\mathcal{C}_2,\mathcal{C}_1,\mathcal{C}_0]\cdot A_9$ and $C$ is monomially equivalent to the matrix-product code
$[\mathcal{C}_8,\ldots,\mathcal{C}_1,\mathcal{C}_0]\cdot A_9$,
where each $\mathcal{C}_\rho$ is a cyclic code of length $10$ over $R$ given by

\par
  $\mathcal{C}_8=\mathcal{C}_7=\langle g_8(x)+ug_{9+8}(x)\rangle=\langle f_3(x)f_4(x)+uf_3(x)\rangle$,

\par
  $\mathcal{C}_6=\langle g_6(x)+ug_{9+6}(x)\rangle=\langle f_1(x)f_3(x)f_4(x)+uf_3(x)\rangle$,

\par
  $\mathcal{C}_5=\mathcal{C}_4=\mathcal{C}_3=\mathcal{C}_2=\langle g_5(x)+ug_{9+5}(x)\rangle=\langle f_1(x)f_3(x)f_4(x)+uf_3(x)f_4(x)\rangle$,

\par
  $\mathcal{C}_1=\mathcal{C}_0=\langle g_1(x)+ug_{9+1}(x)\rangle=\langle uf_3(x)f_4(x)\rangle$

\noindent
 and $A_9=A_3\otimes A_3=\left[\begin{array}{ccc} A_3 & A_3 & A_3\cr
 2A_3 & A_3 & 0 \cr A_3 & 0 & 0\end{array}\right]$ (mod $3$) with $A_3=\left[\begin{array}{ccc} 1 & 1 & 1\cr
 2 & 1 & 0 \cr 1 & 0 & 0\end{array}\right]$.
   Let $d_i$ be the minimum Hamming distance of the cyclic code $\mathcal{C}_i$
with length $10$ over $R$. Then $d_8=d_7=2$, $d_6=2$, $d_2=d_3=d_4=d_5=5$ and $d_1=d_0=5$.

\par
  $\bullet$ By Theorem 4.2, $C$ is monomially equivalent to the matrix-product code $[C^{(2)},C^{(1)},C^{(0)}]\cdot A_3$ where
$C^{(j)}=\langle G^{(j)}(x)\rangle$ is a $(1+u)$-constacyclic code of length $30$ over $R$, i.e. an ideal of the ring
$R[x]/\langle x^{30}-(1+u)\rangle$, generated by the polynomial
$G^{(j)}(x)=\prod_{\lambda=0,1}\prod_{l=0}^{2}
\prod_{i_t> 9\lambda +3j+l, \ 1\leq t\leq 4}f_{t}(x)\in\mathbb{F}_3[x]$. Specifically, we have

\begin{eqnarray*}
G^{(2)}(x)&=&\prod_{i_t>6, \ 1\leq t\leq 4}f_t(x)\prod_{i_t>7, \ 1\leq t\leq 4}f_t(x)\prod_{i_t>8, \ 1\leq t\leq 4}f_t(x)\\
 &&\cdot \prod_{i_t>15, \ 1\leq t\leq 4}f_t(x)\prod_{i_t>16, \ 1\leq t\leq 4}f_t(x)\prod_{i_t>17, \ 1\leq t\leq 4}f_t(x)\\
 &=&f_1(x)f_3(x)f_4(x)\cdot f_3(x)f_4(x)\cdot f_3(x)f_4(x) \cdot f_3(x)^3\\
 &=&f_1(x)f_3(x)^6f_4(x)^3,
\end{eqnarray*}
\begin{eqnarray*}
G^{(1)}(x)&=&\prod_{i_t>3, \ 1\leq t\leq 4}f_t(x)\prod_{i_t>4, \ 1\leq t\leq 4}f_t(x)\prod_{i_t>5, \ 1\leq t\leq 4}f_t(x)\\
 &&\cdot \prod_{i_t>12, \ 1\leq t\leq 4}f_t(x)\prod_{i_t>13, \ 1\leq t\leq 4}f_t(x)\prod_{i_t>14, \ 1\leq t\leq 4}f_t(x)\\
 &=&(f_1(x)f_3(x)f_4(x))^3\cdot (f_3(x)f_4(x))^3\\
 &=&f_1(x)^3f_3(x)^6f_4(x)^6,
\end{eqnarray*}
\begin{eqnarray*}
G^{(0)}(x)&=&\prod_{i_t>0, \ 1\leq t\leq 4}f_t(x)\prod_{i_t>1, \ 1\leq t\leq 4}f_t(x)\prod_{i_t>2, \ 1\leq t\leq 4}f_t(x)\\
 &&\cdot \prod_{i_t>9, \ 1\leq t\leq 4}f_t(x)\prod_{i_t>10, \ 1\leq t\leq 4}f_t(x)\prod_{i_t>11, \ 1\leq t\leq 4}f_t(x)\\
 &=&(f_1(x)f_2(x)f_3(x)f_4(x))^2\cdot f_1(x)f_3(x)f_4(x)\cdot (f_3(x)f_4(x))^3\\
 &=&f_1(x)^3f_2(x)^2f_3(x)^6f_4(x)^6.
\end{eqnarray*}
Hence $C^{(2)}=\langle f_1(x)f_3(x)^6f_4(x)^3\rangle$, $C^{(1)}=\langle f_1(x)^3f_3(x)^6f_4(x)^6\rangle$ and
 $C^{(0)}=\langle f_1(x)^3f_2(x)^2f_3(x)^6f_4(x)^6\rangle$.  By Lemma 3.7, we have
$|C^{(2)}|=3^{60-(1+4\cdot 6+4\cdot 3)}=3^{23}$, $|C^{(1)}|=3^{60-(3+4\cdot 6+4\cdot 6)}=3^{9}$,
$|C^{(0)}|=3^{60-(3+2+4\cdot 6+4\cdot 6)}=3^{7}$. Moreover, from the proof of Theorem 4.2 we
deduce the following conclusions:

\par
  $\diamond$ $C^{(2)}$ is monomially equivalent to $[\mathcal{C}_8,\mathcal{C}_7,\mathcal{C}_6]\cdot A_3$.

\par
  $\diamond$ $C^{(1)}$ is monomially equivalent to $[\mathcal{C}_5,\mathcal{C}_4,\mathcal{C}_3]\cdot A_3$.

\par
  $\diamond$ $C^{(0)}$ is monomially equivalent to $[\mathcal{C}_2,\mathcal{C}_1,\mathcal{C}_0]\cdot A_3$.

\par
Let $d^{(j)}$ be the minimum Hamming distance of
$C^{(j)}$ for $j=0,1,2$.
Since $A_3$ is NSC,
 by Theorems 4.2 and 2.1 it follows that
$d^{(2)}={\rm min}\{3d_8, 2d_7,d_6\}=2$,
$d^{(1)}={\rm min}\{3d_5, 2d_4,d_3\}=5$ and
$d^{(0)}={\rm min}\{3d_2, 2d_1,d_0\}=5$.

\par
  By Theorem 2.1, the minimum Hamming distance of $C=C_{(7,2,18,15)}$ is equal to
$d={\rm min}\{3d^{(2)},2d^{(1)},d^{(0)}\}=5$.


\section{Conclusion} \label{}
\noindent
For any positive integers $m,e$ and a prime number $p$,  denote $R=\mathbb{F}_{p^m}[u]/\langle u^e\rangle$ which is a finite chain ring.
Let $\omega\in R^\times$, $k$ and $n$ be positive integers satisfying ${\rm gcd}(p,n)=1$. We prove that
any $(1+\omega u)$-constacyclic code of length $p^kn$ over $R$ is monomially equivalent to a matrix-product code of a nested sequence of $p^k$ cyclic codes with length $n$ over $R$ and a $p^k\times p^k$ matrix $A_{p^k}$ over $\mathbb{F}_p$. Then we give an iterative construction of every $(1+\omega u)$-constacyclic code by $(1+\omega u)$-constacyclic codes of shorter lengths over $R$.
The next work is to rediscover new properties
for minimum distance of the codes by use of their
matrix-product structures.

\vskip 3mm \noindent {\bf Acknowledgments}
 Part of this work was done when Yonglin Cao was visiting Chern Institute of Mathematics, Nankai University, Tianjin, China. Yonglin Cao would like to thank the institution for the kind hospitality. This research is
supported in part by the National Natural Science Foundation of
China (Grant Nos. 11671235, 61571243, 11471255).





\end{document}